\documentclass[pjabdraft]{pjab}
\usepackage{textcomp}
\usepackage{amsmath,amsfonts,amssymb,mathrsfs,bm,bbm}
\usepackage[dvipdfmx]{graphicx}
\usepackage{array}
\usepackage[authoryear]{natbib}

\usepackage[dvipdfmx]{graphics} 
\usepackage{xcolor}
\usepackage{url}
\usepackage{hyperref}
\usepackage[OT2,T1]{fontenc}
 \definecolor{applegreen}{rgb}{0.55,0.71,0.0}
 \definecolor{forestgreen}{rgb}{0.13,0.55,0.13}
 \definecolor{pinegreen}{rgb}{0.0,0.47,0.44}
 \definecolor{upforestgreen}{rgb}{0.0,0.27,0.13}
 \definecolor{vividviolet}{rgb}{0.62,0.0,1.0}
 \definecolor{lightcyan}{rgb}{0.717647, 0.933071, 0.996078} % in RGB
 \definecolor{scarlet}{rgb}{1.0,0.13,0.0}

\usepackage[yyyymmdd,hhmmss]{datetime}

\usepackage{bm}
\usepackage{ulem}
\DeclareRobustCommand{\erase}{\bgroup\markoverwith{\textcolor{red}{\rule[.5ex]{2pt}{1.2pt}}}\ULon}

\usepackage{soul}

 \definecolor{applegreen}{rgb}{0.55,0.71,0.0}
 \definecolor{forestgreen}{rgb}{0.13,0.55,0.13}
 \definecolor{pinegreen}{rgb}{0.0,0.47,0.44}
 \definecolor{upforestgreen}{rgb}{0.0,0.27,0.13}
 \definecolor{vividviolet}{rgb}{0.62,0.0,1.0}
 \definecolor{lightcyan}{rgb}{0.717647, 0.933071, 0.996078} % in RGB
 \definecolor{scarlet}{rgb}{0.7421875, 0.00390625, 0.09765625}
 \definecolor{darkred}{rgb}{.54296875, 0, 0}

 \def\myfigwidthF{1.00\textwidth}
 \def\myfigwidthM{0.8\textwidth}

\makeatletter
\def\DD{\@ifnextchar [{\D@D}{\D@D[]}}
\def\D@D[#1]#2#3{\frac{d^{#1} #2}{d #3{}^{#1}}}
\def\DP{\@ifnextchar [{\D@P}{\D@P[]}}
\def\D@P[#1]{\@ifnextchar [{\D@Pi[#1]}{\D@PD[#1]}}
\def\D@PD[#1]#2#3{\frac{\partial^{#1} #2}{\partial #3{}^{#1}}}
\def\D@Pi[#1][#2]#3#4{\left( \frac{\partial^{#1} #3}{\partial #4{}^{#1}}\right)_{#2}}
\makeatother

 \newcommand{\mysymfigO}{{Fig. }}
 \def\mtxtsf#1{{\textsf{#1}}}

 \newcommand{\Oalsqr}{O\left(\alpha^2\right)}
\newcommand{\tpspcD}{\;\;}

 \def\mainword{oscillation}
\begin{document}

% Definition for Babel
%\selectlanguage{english}

% Put line numbers
%\linenumbers
\nolinenumbers  % for arXiv (main text)

%%% The PJA style
\Vol{100}
\No{XXX}
%\pjabcategory{Review}
 \pjabcategory{Original Article}
\title[Yoshihide Kozai and his life]
%     {Re-recognized Universality of Yoshihide Kozai's work on three-body dynamics}
      {Re-recognized Universality of Kozai Oscillation on Three-Body Dynamics}
%\subtitle{subtitle}
\authorlist{%
 \Cauthorentry{Masanori Iye}{labelA,labelB}% <- corresponding author
  \authorentry{Takashi Ito}{labelA,labelC,labelD}
 %\authorentry{Sei Mei}{labelA,LabelB}
}
\affiliate[labelA]{National Astronomical Observatory of Japan, Osawa 2--21--1, Mitaka 181--8588, Tokyo, Japan}
\affiliate[labelB]{The Japan Academy, 7--32 Ueno Park,                      Taito--ku 110--0007, Tokyo, Japan}
\affiliate[labelC]{PERC, Chiba Institute of Technology, 2--17--1 Tsudanuma, Narashino 275--0016, Chiba, Japan}
\affiliate[labelD]{Chubu University, 1200 Matsumoto--cho,                     Kasugai 487--8501, Aichi, Japan}
\Correspondence{Masanori Iye (\textsf{masanori.iye2@gmail.com})}
%\Abbreviations{PJA: Proceedings of the Japan Academy; RNA: Ribonucleic acid}

\abstract{%
\nolinenumbers  % for arXiv (abstract)
In 1962, Yoshihide Kozai reported his findings on the secular dynamics of asteroids moving in orbits with high inclination and eccentricity.
In contrast to the classic understanding of the stability of planetary motion in the solar system, Kozai showed that asteroids can significantly change their orbital shape over a long timescale in an oscillatory manner between nearly circular orbits and highly elliptic orbits.
An anti-correlated variation between orbital inclination and eccentricity characterizes this oscillation.
The importance of Kozai's work in understanding the dynamical evolution of various systems was recognized decades later, including the fields of irregular satellites of planets, Oort Cloud, extrasolar planets, binary star systems, type Ia supernovae, planet climate, merging black hole systems, and so on.
}

%%\ed-mja{Edited by xxx \textsc{yyy}, ...}
%%\received{2021}{7}{28}
%%\accepted{2021}{8}{31}
%%\doi{10.2183}{96.014}

\renewcommand{\baselinestretch}{1.200} 
%\begin{center}
%  \textcolor{red}{\LARGE \textsf{Last updated: {\today} {\currenttime} JST}}
%\end{center}

\maketitle

\clearpage
\renewcommand{\baselinestretch}{1.020}
\renewcommand{\baselinestretch}{1.655}  % For the draft mode
\clearpage

In this paper, we review Yoshihide Kozai's work on three-body dynamics and its increasing impact on various fields of astronomy.
We describe the perturbation theory of the three-body problem in Section 1 and summarize the essence of Kozai's work so that readers can trace the mathematical procedures in a comprehensive and self-contained way in Section 2.
Section 3 presents examples where Kozai's work has recently been highlighted.
Section 4 describes his academic life.
We have added a discussion in Section 5.

For this article, we consulted two standard textbooks \citep{brouwer1961,murray1999} and a monograph \citep{ito2019} among many past studies in this field.
\footnote{For readers who understand Japanese, additional relevant articles are available \citep[e.g.][]{kinoshita2000e,kokubo2003}}.

Gravity is one of the most fundamental forces in the universe that governs the motion of celestial bodies.
It is always a force of attraction and works over an infinite distance.
In this paper, we focus on the motion of objects under Newtonian dynamics and do not consider relativistic effects.

\section{Perturbation theory in celestial mechanics}
%%%%%%%%%%%%%%%%%%%%%%%%%%%%%%%%%%%%%%%%%%%%%%%%%%%%%%%%%%20250116mornig
\subsection{Two-body problem\label{ssec:two-body}}
We begin by considering the case where two objects exert gravity on each other.
This circumstance is called the gravitational two-body problem or simply the two-body problem.

Suppose the two objects' masses are $M$ and $m$, respectively.
The equations of motion for each object can be described, i.e., by two sets of ordinary differential equations.
However, we can reduce the two-body problem to the relative motion of a single body by describing the system on the coordinates concerning one of the objects.
In this case, we can consider only one equation of motion that describes the relative motion.

Assume $\bm{r}$ is the position vector from mass point $M$ to mass point $m$, and $t$ is time.
Then, the equation for the relative motion of mass $m$ with respect to mass $M$ is as follows:
\begin{equation}
 \DD[2]{\bm{r}}{t} = - \mu \frac{\bm{r}}{r^3},
 \label{eqn:eom-twobody}
\end{equation}
where
\begin{equation}
 \mu = {\cal G} (M + m),
 \label{eqn:def-mu}
\end{equation}
and ${\cal G}$ as the gravitational constant.
The right-hand side of Eq. \eqref{eqn:eom-twobody} denotes the mutual gravity working between the mass points $M$ and $m$.

The solution of the equations of motion of the two-body system \eqref{eqn:eom-twobody} is the well-known Keplerian motion (or the Keplerian orbit).
The Keplerian motion follows the three laws of Kepler, empirically formulated in the early 17th century by analyzing the positions of the solar system planets observed by Tycho Brahe over decades.
There are three types of shapes in the orbits of Keplerian motion: ellipses (including the circle), parabolas, and hyperbolas.
The object's orbital motion is around the other object located at a focus of these quadratic curves.

%%%%%%%%%%%%%%%%%%%%%%
\begin{figure}[!htbp]
\begin{center}
\includegraphics[width=\myfigwidthM]{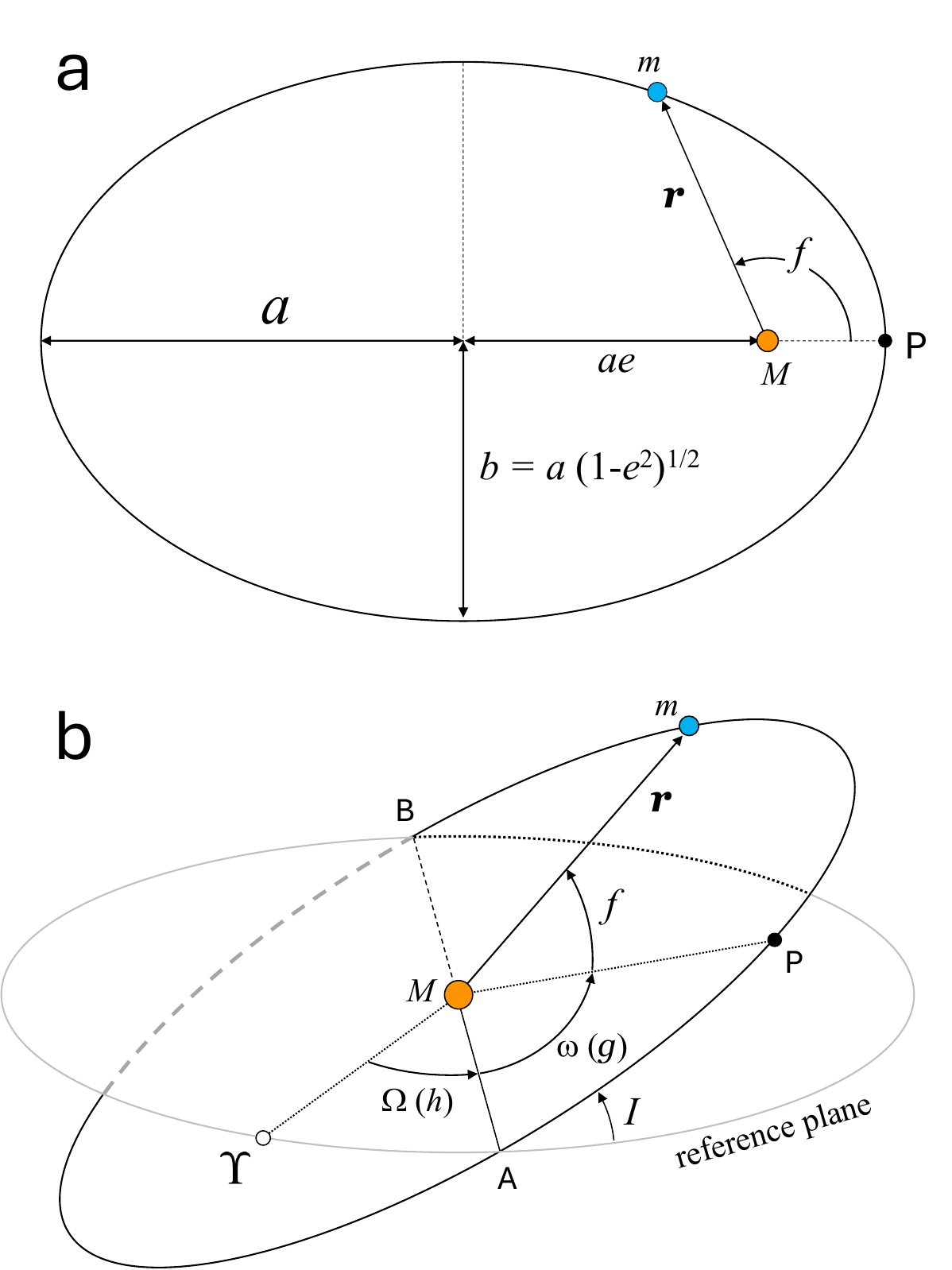}
\end{center}
\caption{%
Keplerian orbital elements that describe an elliptic orbit.
The panel \textsf{a} explains the elements in the orbital plane.
$a$ is the semimajor axis and $b$ is the semiminor axis defined by $b = a \sqrt{1-e^2}$ using eccentricity $e$.
$f$ denotes an angle referred to as the true anomaly.
$\bm{r}$ is the position vector from mass $M$ to mass $m$.
The point \textsf{P} denotes the location of the pericenter.
In this panel, the mass point $M$ is located at one of the ellipse foci.
The distance between the ellipse's center and the foci is $ae$.
The panel \textsf{b} explains the elements concerning the reference plane.
$\Omega$ (or $h$) denotes the longitude of the ascending node, and
$\omega$ (or $g$) denotes the argument of the pericenter.
In this panel, the mass point $M$ is still located in one of the foci and not in the ellipse's center.
The point \textsf{A} denotes the location of the ascending node and
          \textsf{B} denotes the location of the descending node.
We assume that mass $m$ orbits counterclockwise in this figure.
$\Upsilon$ denotes a reference direction fixed in the inertial space.
The reference plane is also fixed in the inertial space.
}
\label{fig:orbit2}
\end{figure}

Six variables, called orbital elements, define the Keplerian motion.
Here is a brief description of the orbital elements, particularly for the elliptic orbits.
Consult Figure \ref{fig:orbit2} for the geometric relationship of each element (except for the sixth time-dependent one).
\begin{itemize}
\item Semimajor axis $a$ \\
This element describes the size of the orbit.
It represents the longer radius of the ellipse, i.e., half of its longest diameter.
This quantity is directly related to the orbital period of Keplerian motion.
If we write the orbital period as $P$ and define the mean angular velocity (the mean motion) $n$ as follows:
\begin{equation}
  n = \frac{2\pi}{P},
  \label{eqn:def-meanmotion}
\end{equation}
Then, Kepler's third law states the following:
\begin{equation}
  n^2 a^3 = \mu .
  \label{eqn:KeplersLaw-3}
\end{equation}
The semimajor axis is also directly related to the orbital energy of the two-body system.
The total orbital energy (the sum of kinetic energy and potential energy) is given as follows:
\begin{equation}
  E = -\frac{\mu}{2a} .
  \label{eqn:orbitalenergy-by-a}
\end{equation}
\item Eccentricity $e$ \\
This element describes the orbit's shape and indicates the degree of deviation from a complete circle.
The closer $e$ is to 0, the more circular the orbit is.
As $e$ approaches 1, the ellipse becomes more elongated.
When $e = 1$, the orbit is a parabola.
When $e > 1$, the orbit is a hyperbola.
When $e \geq 1$, the orbit is not closed, and the motion is not periodic.
The eccentricity $e$ is connected to the angular momentum of the two-body system and is proportional to the quantity $\sqrt{\mu a ( 1-e^2)}$.
\item Inclination $I$  \\
This element is the angle between the Keplerian orbit plane and the reference plane.
For planets in the solar system, the reference plane is typically the ecliptic plane (Earth's orbital plane around the Sun) at a given epoch.
Inclination $I$ is measured in radians, ranging from 0 to $\pi$.
When $I < \frac{\pi}{2}$, the orbital motion is called prograde.
When $I > \frac{\pi}{2}$, the orbital motion is called retrograde.
When $I = \frac{\pi}{2}$, it is called the polar orbit.
The inclination $I$ is related to the vertical component of the angular momentum vector of the two-body system and is proportional to the quantity $\sqrt{\mu a ( 1-e^2)} \cos I$.
\item Longitude of ascending node $\Omega$ (or $h$) \\
This element describes the orientation of the Keplerian orbit in three-dimensional space.
The range of this element is from 0 to $2 \pi$.
This defines the angle between a reference direction (usually the direction of the vernal equinox $\Upsilon$ fixed in inertial space) and the orbit's ascending node.
The ascending node is where the orbiting body crosses the reference plane from south to north.
\item Argument of pericenter $\omega$ (or $g$) \\
This element describes the orientation of the orbit within its orbital plane.
The range of this element is from 0 to $2 \pi$.
Specifically, it is the angle between the ascending node of the orbit and the point closest to the central mass (the pericenter or periastron).
When the central mass is the Sun or Earth, this is called the argument of perihelion or the argument of perigee, respectively.
\end{itemize}

The above five variables are time-independent constants for Keplerian orbits. The remaining sixth orbital element is time-varying.
\begin{itemize}
\item A variable directly connected to time \\
The true anomaly, the angle $f$ in Figure \ref{fig:orbit2}, is the angular distance from the pericenter to the object.
Its rate of change is not proportional to time unless eccentricity $e$ is zero.
The most commonly used element directly related to time $t$ is the mean anomaly $l$, which is defined by:
\begin{equation}
  l = n (t - t_0),
  \label{eqn:def-meananomaly}
\end{equation}
where $n$ is the mean motion defined by Eq. \eqref{eqn:def-meanmotion}  and it is a constant.
$t_0$ is when $l=0$ is realized, i.e., the time the object passes its pericenter.
The mean anomaly $l$ is connected to an angle called the eccentric anomaly $u$ as follows:
\begin{equation}
  l = u - e \sin u .
  \label{eqn:KeplerEquation}
\end{equation}
The eccentric anomaly $u$ is connected to the true anomaly $f$ through the geometric relationship shown in Figure \ref{fig:def-eccanom}.
For example, the following relationships are often useful \citep[e.g.][their p.~24--25]{brouwer1961}:
\begin{equation}
  \tan \frac{f}{2} = \sqrt{\frac{1+e}{1-e}} \tan \frac{u}{2} ,
  \label{eqn:tanf2-tanu2}
\end{equation}
and
\begin{equation}
  \cos f = \frac{\cos u - e}{1 - e \cos u}, \quad
  \sin f = \frac{\sqrt{1-e^2}\sin u}{1 - e \cos u} .
\label{eqn:cosf-sinf-by-u}
\end{equation} 

%%%%%%%%%%%%%%%%%%%%%%
\begin{figure}[!htbp]
\begin{center}
\includegraphics[width=\myfigwidthM]{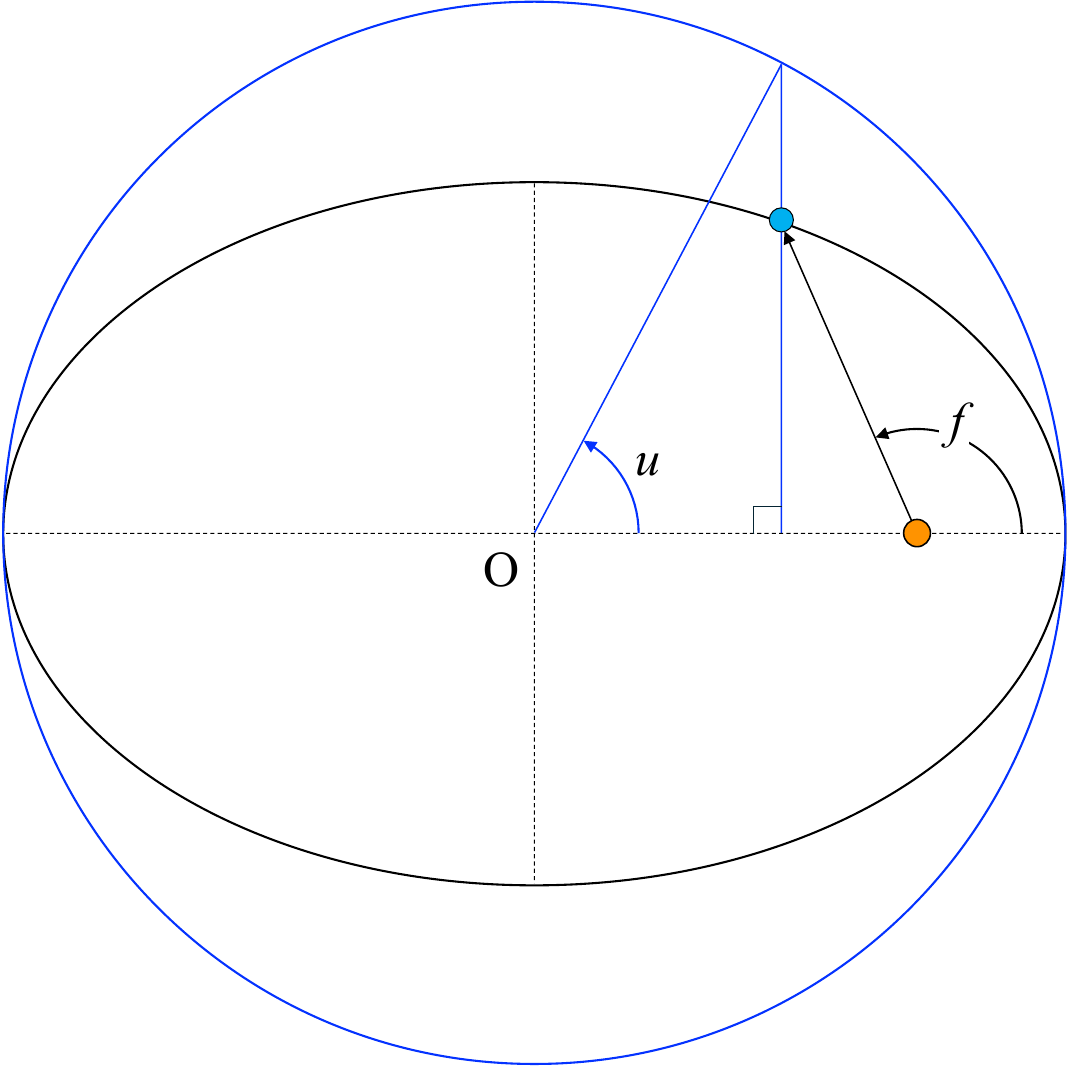}
\end{center}
\caption{%
Geometric relationship between eccentric anomaly $u$ and true anomaly $f$.
Here, O is the ellipse's center (the black curve).
The outer blue circle is the circumscribed circle of the orbital ellipse, within which the eccentric anomaly is defined.
}
\label{fig:def-eccanom}
\end{figure}

The equation \eqref{eqn:KeplerEquation} is called Kepler's equation, and it links the position and time of a celestial object in its orbit.
However, Kepler's equation is transcendental and cannot be solved algebraically for $u$ as $u = u(l)$.
Instead, numerical methods (iterative techniques) or polynomial approximations are typically used to find the solution of $u=u(l)$ for a given $l$.
\citet[][his Section 6.6]{danby1992} explains in detail how to employ numerical methods to solve Kepler's equation in an accessible manner for beginning learners.

The distance between the two bodies in the Keplerian orbit $r$ is described using true anomaly $f$ as follows:
\begin{equation}
  r = \frac{a \left( 1-e^2 \right)}{1 + e \cos{f}} .
  \label{eqn:r-by-f}
\end{equation}
It can also be expressed as follows using eccentric anomaly $u$:
\begin{equation}
  r = a (1 - e\cos u) .
\label{eqn:r-by-u}
\end{equation}

\end{itemize}

For more detailed discussions on the general principles of the two-bode problem as described in this section, readers should consult standard, classical textbooks such as
\citet{smart1953},
\citet{brouwer1961},
\citet{danby1992},
\citet{boccaletti1996,boccaletti1998},
\citet{murray1999},
\citet{merritt2013}, or
\citet{tremaine2023}.

%%%%%%%%%%%%%%%%%%%%%%%%%%%%%%%%%%%%%%%%%%%%%%%%%%%%%%%%%%%%%%%
\subsection{Three-body problem}

%%%%%%%%%%%%%%%%%%%%%%
\begin{figure}[!htbp]
\begin{center}
\includegraphics[width=\myfigwidthF]{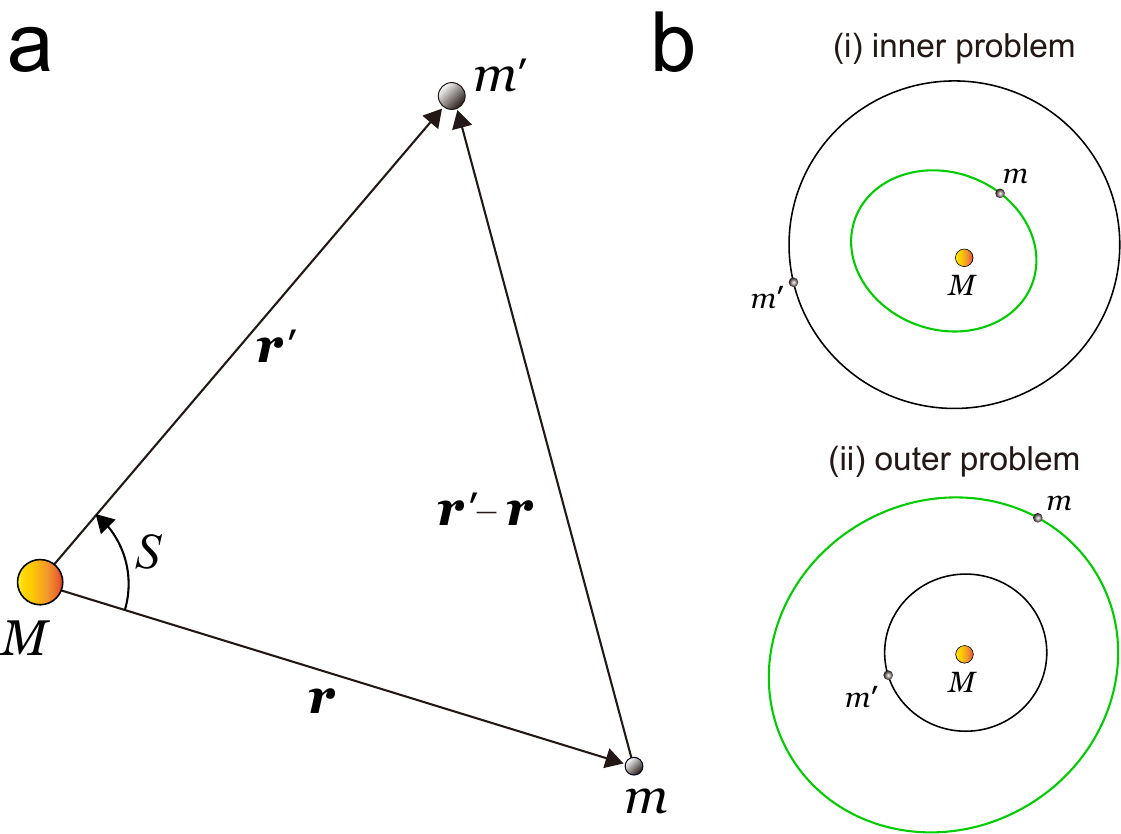}
\end{center}
\caption{%
Schematic illustrations of the relative motion of the hierarchical three-body system.  \mtxtsf{a}: Relative geometric configuration of a three-body system $(M, m', m)$ with respect to the primary mass $M$.
 The vectors $\bm{r}$ and $\bm{r}'$ are position vectors from the mass $M$ to the masses $m$ and $m'$, respectively.
  We presume that the masses have the relation of $M > m' > m$ here.
  \mtxtsf{b}: Two typical patterns of the restricted three-body problem.
\mtxtsf{(i)} The inner problem where the orbit of the perturbed body $m$ (the green ellipse) lies inside the orbit of the perturbing body $m'$ (the black ellipse).
\mtxtsf{(ii)}  The outer problem where the perturbed body's orbit (the green ellipse) lies outside the orbit of the perturbing body (the black ellipse).
This figure is modified and adopted from \citet[][their Fig. 1]{ito2019}.
} 
\label{fig:relativemotion}
\end{figure}

The two-body problem is integrable.
It has an exact analytic solution, the Keplerian motion.
However, by just adding one additional object to the system, the system exhibits quite a different behavior:
The three-body problem is not integrable \citep[e.g.][their Section 3]{ito2007c}, and we have no exact analytic solution except in very few exceptional cases \citep{bruns1887,poincare1890}.

Although the three-body problem is generally non-integrable, the nearly integrable hierarchical three-body system described below serves as a valuable platform for studying the motion of objects in the solar system \citep[e.g.][]{szebehely1967}.
A hierarchical three-body system comprises
a massive central primary mass $(M)$ accompanied by
a smaller secondary mass $(m' < M)$, as well as
an even smaller tertiary mass $(m < m')$.
\mysymfigO \ref{fig:relativemotion}\mtxtsf{a} schematically shows the relative locations of the three masses.
The tertiary mass can orbit inside or outside the $(M,m')$ binary as shown in \mysymfigO \ref{fig:relativemotion}\mtxtsf{b}.
Unless the orbits of the secondary and tertiary around the primary get too close or intersect each other, the two binaries, $(M,m')$ and $(M,m)$, usually both behave almost like the Keplerian motions.
In this case, we can obtain their orbital solution primarily through perturbation methods \citep[e.g.,][]{hori1966}.
The binary $(M,m')$ would make a pure Keplerian motion if $m \ll m'$, not being disturbed by $m$ at all, while the motion of $m$ is affected by the binary $(M,m')$.
This is called the restricted three-body problem (denoted hereafter R3BP).
In particular, when the orbit of $m'$ in the binary $(M,m')$ is circular, the system results in the circular restricted three-body problem (CR3BP).
This is the problem \citet{kozai1962b} intended to study.

%%%%%%%%%%%%%%%%%%%%%%%%%%%%%%%%%%%%%%%%%%%%%%%%%%%%%%%%%%
\subsection{Disturbing function\label{ssec:CR3BP-R}}
Following the derivation described in standard textbooks \citep[e.g.][]{brouwer1961,danby1992,murray1999}, we have the equation of motion of the mass point $m$ relative to the central mass $M$ as follows:
\begin{equation}
  \DD[2]{\bm{r}}{t}  + \mu\frac{\bm{r}}{r^3}   = \nabla R .
  \label{eqn:eom-relative-useR-conventional-ndash}
\end{equation}

On the other hand, the equation of motion of the mass point $m'$ relative to the central mass $M$ is expressed as follows:
\begin{equation}
  \DD[2]{\bm{r}'}{t} + \widetilde{\mu} \frac{\bm{r}'}{{r'}^3} = \nabla R' ,
  \label{eqn:eom-relative-useR-conventional-wdash}
\end{equation}
where
\begin{equation}
 \widetilde{\mu} = {\cal G} (M + m') .
 \label{eqn:def-tildemu} 
\end{equation}

$R$ in Eq. \eqref{eqn:eom-relative-useR-conventional-ndash} as well as
$R'$ in Eq. \eqref{eqn:eom-relative-useR-conventional-wdash} are called disturbing functions.
The function forms are as follows:
\begin{equation}
  R  = \frac{{\cal G} m'}{\Delta}
             -{\cal G} m'  \frac{\bm{r}  \cdot \bm{r}'}{{r'}^3} ,
  \label{eqn:def-R-conventional-ndash}
\end{equation}
and
\begin{equation}
  R'
     = \frac{{\cal G} m }{\Delta}
            -{\cal G} m   \frac{\bm{r}' \cdot \bm{r}}{r^3} ,
  \label{eqn:def-R-conventional-wdash}
\end{equation}
with the conventional notation for mutual distance
\begin{equation}
  \Delta \equiv \left| \bm{r}' - \bm{r}  \right|
         =      \left| \bm{r}  - \bm{r}' \right| .
  \label{eqn:def-Delta-0}
\end{equation}

The disturbing functions represent the gravitational interaction between mass $m$ and mass $m'$.
The first term of the disturbing function \eqref{eqn:def-R-conventional-ndash} or \eqref{eqn:def-R-conventional-wdash} is called the direct part and represents the major component of the mutual perturbation between the two masses.
The second term is called the indirect part.
The indirect part vanishes or becomes constant when we average the system over the orbit \citep[e.g.][]{murray1999}.

We expand the disturbing function, $R$ or $R'$, in an infinite series of orbital elements using Legendre polynomials.
Applying the cosine formula to the triangle $m$--$M$--$m'$ with angle $S$
in \mysymfigO \ref{fig:relativemotion}\mtxtsf{a}, we obtain
\begin{equation}
  \left| \bm{r}' - \bm{r}  \right|^2 = r^2 + {r'}^2 - 2 r r' \cos S .
  \label{eqn:cosineformula}
\end{equation}
Using Eq. \eqref{eqn:cosineformula}, we can expand the inverse of $\Delta$ in Eq. \eqref{eqn:def-Delta-0} through the Legendre polynomials $P_j (\cos S)$.
When $r < r'$, the expansion becomes
\begin{equation}
\begin{aligned}
  \frac{1}{\Delta} &= \frac{1}{r'}
    \left( 1 - 2 \frac{r}{r'} \cos S + \left( \frac{r}{r'} \right)^2 \right)^{-\frac{1}{2}} \\
  & = \frac{1}{r'}
    \sum_{j=0}^{\infty} \left( \frac{r}{r'} \right)^j P_j (\cos S) .
\end{aligned}
  \label{eqn:Delta-expansion-Legendre-inner}
\end{equation}
On the other hand, when $r > r'$, it becomes
\begin{equation}
\begin{aligned}
  \frac{1}{\Delta} &= \frac{1}{r}
    \left( 1 - 2 \frac{r'}{r} \cos S + \left( \frac{r'}{r} \right)^2 \right)^{-\frac{1}{2}} \\
  & = \frac{1}{r}
    \sum_{j=0}^{\infty} \left( \frac{r'}{r} \right)^j P_j (\cos S) .
\end{aligned}
  \label{eqn:Delta-expansion-Legendre-outer}
\end{equation}

In the remainder of this section, we discuss the restricted three-body problem (R3BP) and assume $m \ll m'$.
Let us remember here that $m$ is the mass of the perturbed object, and $m'$ is the mass of the perturbing object.
The orbital condition $r<r'$ required for the expression of $1/\Delta$ in Eq. \eqref{eqn:Delta-expansion-Legendre-inner} indicates that we are dealing with the inner problem
(see \mysymfigO \ref{fig:relativemotion}\mtxtsf{b(i)}).
In this case, the term with $j=0$ in Eq. \eqref{eqn:Delta-expansion-Legendre-inner} does not depend on $r$ and disappears through its differentiation \citep[e.g.][their Section 6.3, p. 229]{murray1999}.
Therefore, we can ignore
the $j=0$ terms in Eq. \eqref{eqn:Delta-expansion-Legendre-inner}.
In addition, we have the following relationship,
\begin{equation}
  \bm{r} \cdot \bm{r'} = r r' \cos S = r r' P_1 (\cos S),
  \label{eqn:def-rrdcosS}
\end{equation}
and the indirect part of the disturbing function cancels out the term of $j=1$ in Eq. \eqref{eqn:Delta-expansion-Legendre-inner}.
Therefore, to construct the disturbing function of the inner problem, we consider only the $j \geq 2$ terms in the expansions of
$1/\Delta$ in Eq. \eqref{eqn:Delta-expansion-Legendre-inner} as
\begin{equation}
\begin{aligned}
  \frac{1}{\Delta}
= \frac{1}{r'}
    \sum_{j=2}^{\infty} \left( \frac{r}{r'} \right)^j P_j (\cos S) .
\end{aligned}
  \label{eqn:Delta-expansion-Legendre-inner-nge2}
\end{equation}

%%%
%%% The outer problem - briefly
%%%
For the outer problem where condition $r>r'$ holds (see \mysymfigO \ref{fig:relativemotion}\mtxtsf{b(ii)}), the expression of $1/\Delta$ in Eq. \eqref{eqn:Delta-expansion-Legendre-outer} should be employed. 
The treatment of the outer problem is more complicated than the inner problem. However, the principle is the same. Readers can consult previous works \citep[e.g.][]{thomas1996,ito2016} or textbooks \citep{murray1999} for more details on the treatment of outer problems.
Yoshihide Kozai also provided a concise review of the treatment of the outer problem and its application to Pluto's motion, publishing it in \textit{the Proceedings of the Japan Academy} \citep{kozai2004}.

In the circular restricted three-body problem (CR3BP), the length of the position vector of the perturber $(r')$ with respect to the primary mass has a constant value equivalent to its semimajor axis $a'$ because the perturber's eccentricity is zero $(e'=0)$.
When $r'=a'$, we do not need to consider odd terms $(j=3,5,7,\cdots)$ in the expansion of
Eqs. \eqref{eqn:Delta-expansion-Legendre-inner} or
     \eqref{eqn:Delta-expansion-Legendre-outer} 
because they vanish after the averaging procedure (see Appendix \ref{appen:Pnodd} for more details).
% \citep[e.g.][]{brouwer1961}.
Therefore, the disturbing function of the inner CR3BP is simplified from
Eqs. \eqref{eqn:def-R-conventional-ndash} and
     \eqref{eqn:Delta-expansion-Legendre-inner} to:
\begin{equation}
\begin{aligned}
  R = \frac{{\cal G} m'}{a'}
    \sum_{n=1}^{\infty} \left( \frac{r}{a'} \right)^{2n} P_{2n} (\cos S) .
\end{aligned}
  \label{eqn:Delta-expansion-Legendre-iCR3BP}
\end{equation}
This is the disturbing function that \citet{kozai1962b} considered.
Section \ref{sec:Kozaiswork} summarizes his discussion based on this function.

Note that in the hierarchical $N$-body problem, including the three-body problem addressed in this paper, a set of canonical coordinates known as the Jacobi coordinates are commonly used \citep[e.g.,][]{plummer1960, wisdom1991}.
In the Jacobi coordinates, the position and velocity of object $k$ are given relative to the barycenter of objects $0$ through $k-1$ \citep[e.g. See Figure 9.20 of][]{murray1999}.
Usually, index 0 is assigned to the central mass (e.g., the Sun), and other indices are assigned to the remaining objects in the order of increasing semimajor axis.
For the Sun--Jupiter--asteroid system, Jupiter's coordinate origin is the Sun, consistent with the relative coordinates employed in this paper.
However, the asteroid's coordinate origin was the barycenter of the Sun--Jupiter binary rather than the Sun itself.
Because the Jacobi coordinates are intrinsically canonical, the Hamiltonian can be described in a more compact form than relative or barycentric coordinates \citep[e.g.][]{tremaine2023}.
In addition, the Jacobi coordinates facilitate the development of a highly efficient algorithm for numerical orbit integration in a weakly perturbed $N$-body system \citep{wisdom1991}.
It is straightforward to show that
% the direct part of the disturbing function $1/\Delta$
$1/\Delta$ included in the direct part of the disturbing function for the inner problem as seen in Eq. \eqref{eqn:Delta-expansion-Legendre-inner} or in Eq. \eqref{eqn:Delta-expansion-Legendre-inner-nge2} can be derived from the general Hamiltonian described using the Jacobi coordinates \citep[][their p.~7]{ito2019}.
However, note again that we do not use the Jacobi coordinates in this paper and instead stay on the relative coordinates, as Kozai employed the relative coordinates in his study.

%%%%%%%%%%%%%%%%%%%%%%%%%%%%%%%%%%%%%%%%%%%%%%%%%%%%%%%%%%
\section{Kozai's work in 1962\label{sec:Kozaiswork}}

The leading publication we introduce here is \citet{kozai1962b}, published in \textit{The Astronomical Journal\/}.
This paper was more widely and quickly recognized than other studies on the same subject, as mentioned later.
\citet{kozai1962b} was selected as one of the 53 ``Selected Fundamental Papers Published this Century in the Astronomical Journal and the Astrophysical Journal'' \citep{abt1999}, and was republished in 1999 on the centennial anniversary of the American Astronomical Society (AAS).
A short editorial by Brian Geoffrey Mardsen, the AAS president at that time, is available \citep{marsden1999}.

In \citet{kozai1962b}, the disturbing function of CR3BP in the Hamiltonian form was first described.
Next, the system's degrees of freedom were reduced using the averaging technique, and the system was brought into an integrable form.
In this section, we first describe his method and then summarize his conclusions.
Additionally, we compare Kozai's calculations with the results of the numerical integration from a modern perspective.

%%%%%%%%%%%%%%%%%%%%%%%%%%%%%%%%%%%%%%%%%%%%%%%%%%%%%%%%%%
\subsection{Averaging of disturbing function}\label{ssec:CR3BP-averaging} 
Having the disturbing function $R$ in the form of Eq. \eqref{eqn:Delta-expansion-Legendre-iCR3BP} in hand, Kozai carried out the double averaging of $R$ by the mean anomalies of the perturbed and perturbing bodies.
This procedure is an important part of reducing the degrees of freedom of the entire system (see Section \ref{ssec:reduction-dof}).
When the motion of an object around the central mass is close to the Keplerian motion, the variation rate of its mean anomaly is orders of magnitude larger than those of the other orbital elements.
Hence, it is justified to eliminate its mean anomaly by averaging, assuming that the other orbital elements of the object do not significantly change over a period of mean anomaly, namely one orbital motion around the central body.
Eliminating the mean anomaly using the averaging procedure can be regarded as part of the canonical transformation that divides the system's Hamiltonian into periodic and secular parts \citep[e.g.][]{boccaletti1998}.
Historically, this procedure was devised by \citet{delaunay1860,delaunay1867}, and substantially developed by \citet{vonzeipel1916a,vonzeipel1916b,vonzeipel1917a,vonzeipel1917b}.
% See 
% \citet[][Notes and References in Chapter XVII, their pp. 591--593]{brouwer1961} or
% \citet[][Subsection 12.4]{goldstein2002} for a more detailed background.

To perform the averaging procedure, we assume no significant resonant relationship exists between the mean motions of the perturbed and perturbing bodies.
Bearing this assumption in mind, we select the $n$-th term of the disturbing function $R$ for the inner problem in Eq. \eqref{eqn:Delta-expansion-Legendre-iCR3BP}, and refer to it as $R_{2n}$.
Defining a mass factor $\mu'$ as
\begin{equation}
  \mu' = {\cal G} m' ,
  \label{eqn:def-mudash}
\end{equation}
and use it for simplicity, we obtain
\begin{equation}
  R_{2n} = \frac{\mu'}{a'} \left( \frac{r}{a'} \right)^{2n} P_{2n} (\cos S) .
  \label{eqn:Rd-inner-j}
\end{equation}

First, we average $R_{2n}$ by the mean anomaly $l'$ of the perturbing body.
Using the symbols $\left< \right.$ and $\left. \right>$ for averaging, it is
\begin{equation}
  \left< R_{2n} \right>_{l'} = 
  \frac{\mu'}{a'} \left( \frac{r}{a'} \right)^{2n} \left< P_{2n} \right>_{l'},
  \label{eqn:Rd-inner-j-averaged1}
\end{equation}
where
\begin{equation}
  \left< P_{2n} \right>_{l'} = \frac{1}{2\pi} \int^{2\pi}_0 P_{2n} (\cos S) dl'.
  \label{eqn:Pj-avr}
\end{equation}

\begin{figure}[!htbp]
\begin{center}
\includegraphics[width=\myfigwidthM]{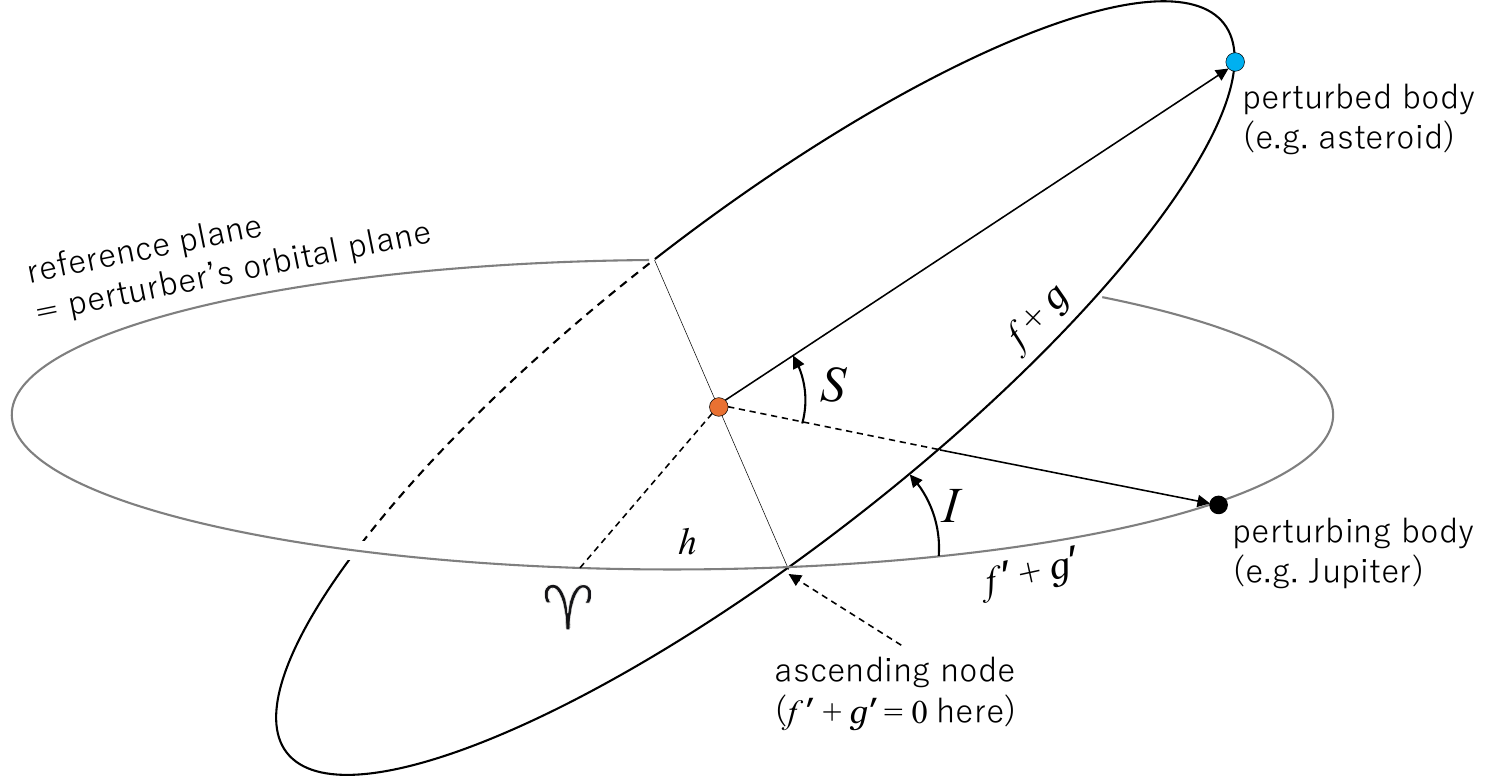}
\end{center}
\caption{%
The spatial configuration of the angle $S$ between perturbed and perturbing bodies in CR3BP.
}
\label{fig:cosS_cr3bp}
\end{figure}

The angle $S$ is expressed by the orbital angles through a spherical trigonometric relationship \citep[e.g.][Eq. (7) on p. 592]{kozai1962b}
\begin{equation}
  \cos S = \cos (f+g) \cos (f'+g')
         + \cos I \sin (f+g) \sin (f'+g'),
  \label{eqn:def-cosS}
\end{equation}
where
$f, f'$ are true anomalies, and $g, g'$ are the arguments of pericenters of the perturbed and perturbing bodies,
respectively, and
$I$ is the mutual inclination angle of the two orbital planes measured at the node of the two orbits.
As for how the angle $S$ is defined in space, we made Figure \ref{fig:cosS_cr3bp} and showed its geometry there.
The orbital plane of the perturbing body was selected as the reference plane to measure $g$ from the mutual node.
The argument of pericenter of the perturbing body $(g')$ is not defined in CR3BP.
Therefore, in Eq. \eqref{eqn:def-cosS}, we regard $f'+g'$ as a single fast-oscillating variable represented by $f'$.
In practice,
we can replace the integral over the mean anomaly of the perturbing body $\int dl'$ for an integral over its true anomaly $\int df'$ in the discussion here because $f' = l'$ in CR3BP. 

%%%%%%%%%%%%%%%%%%%%%%

To obtain $\left< P_{2n} \right>_{l'}$ of Eq. \eqref{eqn:Pj-avr},
we calculate the time average of $\cos^{2n} S$ by $l'$ as follows:
\begin{equation}
    \left< \cos^{2n} S \right>_{l'}
  = \frac{1}{2\pi} \int^{2\pi}_0 \cos^{2n} S dl' .
  \label{eqn:def-cosS-avr}
\end{equation}
Then, we average $\left< R_{2n} \right>_{l'}$ of Eq. \eqref{eqn:Rd-inner-j-averaged1}
by the mean anomaly $l$ of the perturbed bodyas
\begin{equation}
%\begin{aligned}
  \left< \left< R_{2n} \right>_{l'} \right>_{l}
  = \frac{\mu'}{a'} \left( \frac{a}{a'} \right)^{2n}
      \frac{1}{2\pi} \int^{2\pi}_0
        \left( \frac{r}{a} \right)^{2n} \left< P_{2n} \right>_{l'} dl .
%& = \frac{\mu'}{a'} \left( \frac{a}{a'} \right)^{2n} 
%    \frac{1}{2\pi} \int^{2\pi}_0
%          \left( 1-e \cos u \right)^{2n+1} \left< P_{2n} \right>_{l'} du .
% \end{aligned}
\label{eqn:Rd-inner-j-averaged2-byu}
\end{equation}

Eq. \eqref{eqn:Rd-inner-j-averaged2-byu} is the general form of the $n$-th term of the doubly averaged disturbing function for the inner CR3BP.
If we define the ratio of the semimajor axes as $\alpha = \frac{a}{a'} < 1$, this term has the magnitude of $O\left(\alpha^{2n}\right)$.

%%%%%%%%%%%%%%%%%%%%%%%%%%%%%%%%%%%%%%%%%%%%%%
\subsection{Reduction of degrees of freedom\label{ssec:reduction-dof}}

The averaging procedure of the disturbing function described in Section \ref{ssec:CR3BP-averaging} is a part of the procedure to reduce the degrees of freedom of the CR3BP system and transform it into an integrable one.
Thus, we can obtain a comprehensive understanding of the motion of a perturbed object in CR3BP without directly solving the equations of motion.
The following is a summary of \citeauthor{kozai1962b}'s explanation of how the degrees of the disturbing function can be reduced.

In his \citeyear{kozai1962b} paper, \citeauthor{kozai1962b} denoted $m$ as the mass of the asteroid and $m'$ as the mass of Jupiter.
The solar mass is set to unity.
Instead of Kepler orbital elements, \citeauthor{kozai1962b} used the so-called Delaunay elements.
The Delaunay elements are a set of canonical variables that describe the Keplerian motion, and their relationship to the Kepler orbital elements are as follows \citep[e.g.][]{boccaletti1996}:
\begin{equation}
\begin{aligned}
  L &= \sqrt{\mu a},            & \tpspcD l &= \; \mbox{mean anomaly}, \\
  G &= L \sqrt{1-e^2},
                                & \tpspcD g &= \; \mbox{argument of perihelion}\;      (\omega), \\
  H &= G \cos I,                & \tpspcD h &= \; \mbox{longitude of ascending node}\; (\Omega), \\
\end{aligned}
%  \tag{K01-\arabic{equation}}
%  \stepcounter{equation}
  \label{eqn:K01}
\end{equation}
where $l$, $g$, $h$ are the canonical coordinates, and $L$, $G$, $H$ are the corresponding conjugate momenta.
The element $G$ is proportional to the magnitude of the angular momentum vector of the perturbed object (see the description about eccentricity $e$ in Section \ref{ssec:two-body}), and $H$ is proportional to its vertical component (see the description regarding inclination $I$ in Section \ref{ssec:two-body}).

Let us symbolically denote the Hamiltonian $F$ for R3BP using the Delaunay elements as
% $F(L, G, H, l, g, h; L', G,' H,' l,' g,' h')$.
  $F(L, G, H, l, g, h; L', G', H', l', g', h')$.
Here, $L, G, H, l, g, h$ are the canonical variables of the perturbed object, and
% $L', G,' H,' l,' g,' h'$
  $L', G', H', l', g', h'$ are those of the perturbing object.
The Hamiltonian $F$ is the sum of the disturbing part (practically equivalent to the disturbing function $R$) and the orbital energy part (derived from the object's Keplerian motion), and it originally had \it{six} \rm
 degrees of freedom.
However, the Hamiltonian contains $h$ and $h'$ (longitude of ascending node) only in the combination $h-h'$ if we choose the invariable plane as the reference plane.
In other words, the Hamiltonian is symmetric regarding the rotation around the system's total angular momentum vector.
This characteristic is called Jacobi's elimination of the nodes \citep[e.g.][]{jacobi1843a,jacobi1843b,charlier1902,charlier1907} and a typical outcome is $h - h' = \pi$ \citep[e.g.][]{jefferys1966,ford2000}.
This relationship allows us to remove $h$ and $h'$ from the Hamiltonian, and the conjugate momenta $H$ and $H'$ become constants of motion.
Consequently, the original Hamiltonian with six degrees of freedom is converted into $F(L, G, l, g; L', G,' l,' g')$ with {\textit{four} degrees of freedom.

Then, \citeauthor{kozai1962b} averages the Hamiltonian (in other words, disturbing function) using the method described in Section \ref{ssec:CR3BP-averaging} and further reduces its degrees of freedom.
This procedure eliminates mean anomalies of the perturbed body $(l)$ and the perturbing body $(l')$ from the Hamiltonian.
Consequently, their conjugate momenta $L$ and $L'$ become constants of motion, which yields another result through the definition of $L$ and $L'$ in Eq. \eqref{eqn:K01}: the semimajor axes of both objects ($a$ and $a'$) are constant. 
The Hamiltonian with four degrees of freedom is now transformed into a new Hamiltonian $F (G, g; G', g')$ with \textit{two} degrees of freedom.
Here, we recall that \citet{kozai1962b} addressed the CR3BP where the pericenter of the perturbing body is not defined (or $g'$ can be considered constant).
$G' = L'\sqrt{1-{e'}^2}$ is also a constant parameter.
Therefore, we do not need to consider $g'$ or $G'$ in the Hamiltonian, and the function form of the Hamiltonian is $F (G, g)$ with just \textit{one} degree of freedom.}
Because we do not consider energy dissipation, the Hamiltonian $F$ is a constant of motion.
This indicates that any combination of $(G, g)$ uniquely determines the dynamical state of the motion of a perturbed body.
In other words, the doubly averaged CR3BP is integrable.

Recall that the Hamiltonian $F$ can be divided into a Keplerian part (denoted ${\cal K}$) and a perturbation part (denoted $W$). In general, it can be written as follows:
\begin{equation}
  F = {\cal K} + m' W .
  \label{eqn:F_components}
\end{equation}

The factor $m'$ multiplied by $W$ on the right-hand side of Eq. \eqref{eqn:F_components} denotes that the magnitude of the perturbation is proportional to the mass of the perturber.
In general, ${\cal K}$ is a function of one of the Delaunay elements $L = \sqrt{\mu a}$, and is expressed as ${\cal K} \propto -L^{-2}$.
Since the semimajor axis of the perturbed body $a$ is constant after the averaging operation, ${\cal K}$ is also constant.
Therefore, in the following, we discuss only the perturbation part of the Hamiltonian $W$.
Now that we know $W$ depends only on $G$ and $g$, the canonical equations of motion of this system are as follows:
\begin{equation}
  \DD{G}{t} = -m' \DP{W}{g}, \quad
  \DD{g}{t} =  m' \DP{W}{G} .
%  \tag{K10-\arabic{equation}}
%  \stepcounter{equation}
  \label{eqn:K10} % equivalent to Kozai's Eq.(10)
\end{equation}
Due to the traditional sign difference between celestial mechanics and analytic mechanics, $W$ is equivalent to disturbing function $R$ with the sign reversed (i.e., $W = -R$).

Let us note that for the expression of $K$ $(\propto -L^{-2})$ in Eq. \eqref{eqn:F_components}, readers can refer to the expression of $F_0$ on p. 553 of \citet{brouwer1961} except for the sign change from $+$ to $-$.
Additionally, see the expression of $H_0$ in Eq. (10.50) on p. 285 of \citet{boccaletti1998}, with the sign change from $+$ to $-$. 
Note also that the canonical equations of motion \eqref{eqn:K10} can be directly and numerically integrated, and a time-dependent solution can be obtained in the form of $G(t)$ and $g(t)$.
This line of studies using the truncated disturbing function is called semi-analytic, and there is a great deal of literature on it \citep[e.g.][]{laskar1988}.
However, we do not mention it in this paper because Kozai did not do it either.
  
The conservation of one of the Delaunay elements of the perturbed body, $H$, defined in Eq. \eqref{eqn:K01}, plays a crucial role in Kozai's study.
This quantity is proportional to the vertical component of the angular momentum of the perturbed body.
\citeauthor{kozai1962b} introduced a new, constant parameter $\Theta$ which is proportional to the square of $H$ as follows:
\begin{equation}
  \Theta = \left( 1-e^2 \right) \cos^2 I .
  \label{eqn:def-k2}
\end{equation}
The fact that $\Theta$ is a constant means that eccentricity $e$ and inclination $I$ oscillate in an anti-correlated manner.

%%%%%%%%%%%%%%%%%%%%%%%%%%%%%%%%%%%%%%%%%%
\subsection{Stationary point\label{ssec:Kozai-stationary}}
\citet{kozai1962b} then searched for conditions under which the perturbed object's argument of pericenter librates in this Hamiltonian system.
More specifically, he searched for the location of the stationary points that the perturbation part of the doubly averaged disturbing Hamiltonian $W$ can have.
Note that the libration represents an oscillatory motion of an angle around a particular fixed value.
The angle variable in libration does not complete a full circle $(2\pi)$.
On the other hand, circulation is a continuous motion of an angle in one direction that completes a full circle.
At these stationary points, $g$ and $e$ (so $G$) of the perturbed body are supposed to be constant.
To search for stationary points, we must know the specific function form of the disturbing function.
Let us remember that $\alpha$ has been defined as the ratio of the semimajor axis of the perturbed object $a$ and the perturbing object $a'$ as $\alpha = a/a'$.
In the inner CR3BP where $\alpha < 1$,
Kozai selected the terms of the disturbing function in leading order, $\Oalsqr$.
This is equivalent to taking terms up to $2n=2$ from the general form of the disturbing function described in Eq. \eqref{eqn:Rd-inner-j-averaged2-byu}.
Subsequently, the disturbing part of the Hamiltonian $W$ ($= -R$) becomes as follows:
\begin{equation}
 W_{\Oalsqr}
= -\frac{1}{16}
\frac{{\cal G}}{a'} \left( \frac{a}{a'} \right)^2 
\left[
  15 e^2 \sin^2 I \cos 2 g - \left(3e^2+2 \right) \left( 3\sin^2 I -2 \right)
\right].
  \label{eqn:R2-final}
\end{equation}
Readers can refer to Appendix \ref{appen:Wquad} for details on how $W_{\Oalsqr}$ in Eq. \eqref{eqn:R2-final} was derived from Eq. \eqref{eqn:Rd-inner-j-averaged2-byu}.
Note that the truncation of the disturbing function on the order of $\alpha^2$, as in Eq. \eqref{eqn:R2-final}, is commonly referred to as the quadrupole-level (or quadrupole-order) approximation.
In contrast, truncation at the order of $\alpha^3$ is known as the octupole-level (or octupole-order) approximation \citep[e.g.][see her p.~443]{naoz2016}.

Applying $W_{\Oalsqr}$ in Eq. \eqref{eqn:R2-final} to $W$ in Eq. \eqref{eqn:K10}, we can rewrite the canonical equations of motion for $G$ and $g$ as follows:
\begin{alignat}{1}
%%%
  \DD{G}{t}
&=-m' \DP{W_{\Oalsqr}}{g} \nonumber \\
&=   \frac{\mu'}{a'} \left( \frac{a}{a'}\right)^2
     \frac{15}{8} e^2 \sin^2 I \sin 2g, 
\label{eqn:dGdt-quadrupole} \\
  \DD{g}{t}
&= m' \DP{W_{\Oalsqr}}{G} \nonumber \\
&=-\frac{\mu'}{a'} \left( \frac{a}{a'}\right)^2
               \frac{ 3}{8G}
    \left[  \left( 5\cos^2 I - \left(1-e^2\right) \right)
          -5\left(  \cos^2 I - \left(1-e^2\right) \right) \cos 2g \right] .
\label{eqn:dgdt-quadrupole}
\end{alignat}

%\citeauthor{kozai1962b} found that the disturbing Hamiltonian $W$ can have stationary points when $\cos 2 g = -1$.
Let us determine where the stationary points of the disturbing Hamiltonian are located at the quadrupole-level approximation using Eqs. \eqref{eqn:dGdt-quadrupole} and \eqref{eqn:dgdt-quadrupole}.
At the stationary points, both of the following equations must be satisfied:
\begin{equation}
  \DD{G}{t} = 0, \quad
  \DD{g}{t} = 0 .
  \label{eqn:dgdt_dGdt_equalto_0}
\end{equation}

>From Eq. \eqref{eqn:dGdt-quadrupole} and the first equation of Eq. \eqref{eqn:dgdt_dGdt_equalto_0}, we know that the stationary points can be realized when $\sin 2g = 0$.
This means $\cos 2 g = +1$ or $\cos 2 g = -1$.
When $\cos 2 g = +1$, from Eq. \eqref{eqn:dgdt-quadrupole} and the second equation of Eq. \eqref{eqn:dgdt_dGdt_equalto_0} we have
\begin{equation}
  4 \left( 1 - e^2 \right) = 0,
\end{equation}
which yields $e=1$ .
However, we do not consider such a particular case $e=1$.

When $\cos 2 g = -1$, from Eq. \eqref{eqn:dgdt-quadrupole} and the second equation of Eq. \eqref{eqn:dgdt_dGdt_equalto_0} we have
\begin{equation}
  10 \cos^2 I - 6 \left( 1-e^2 \right) = 0 .
  \label{eqn:K-pre26}
\end{equation}

Using Eq. \eqref{eqn:def-k2}, Eq. \eqref{eqn:K-pre26} yields the following relation:
\begin{equation}
  \Theta = \frac{3}{5} \left( 1-e^2 \right)^2 .
  \label{eqn:Theta-otherform1}
\end{equation}

% If we consider elliptic orbits with $0 \leq e < 1$, then we have
As long as we deal with elliptic orbits where $0 \leq e < 1$, we have
\begin{equation}
  0 < \left( 1-e^2 \right)^2 \leq 1 ,
  \label{eqn:range-1e22}
\end{equation}
and thus Eq. \eqref{eqn:Theta-otherform1} gives
\begin{equation}
  \Theta \leq \frac{3}{5} .
  \label{eqn:Theta-range}
\end{equation}

Eq. \eqref{eqn:Theta-range} means that the doubly averaged disturbing Hamiltonian $W_{\Oalsqr}$ in Eq. \eqref{eqn:R2-final} can have stationary points only when the condition \eqref{eqn:Theta-range} is satisfied ---
in other words, the parameter $\Theta$ has an upper limit beyond which the perturbed body's argument of pericenter cannot librate.
Eq. \eqref{eqn:Theta-range}, together with the definition of $\Theta$ in Eq. \eqref{eqn:def-k2}, indicates that there is a lower limit for the perturbed object's inclination for the libration of its argument of pericenter to occur.
We obtain this value by setting $e=0$, and Eq. \eqref{eqn:Theta-range} becomes
\begin{equation}
  \cos^2 I \leq \frac{3}{5} .
  \label{eqn:cos2I-range}
\end{equation}
By assuming $I < 90^\circ$ (the prograde motion of the object), Eq. \eqref{eqn:cos2I-range} can be translated as follows:
\begin{equation}
  I \geq \cos^{-1} \sqrt{\frac{3}{5}} = 39^\circ . 231 \cdots.
  \label{eqn:I-range}
\end{equation}

Eq. \eqref{eqn:I-range} is a representation of the necessary condition for the perturbed object's argument of pericenter to librate at the quadrupole-level approximation --- when the object's inclination $I$ satisfies this condition at $e=0$, the object's argument of pericenter $g$ can librate.

Note that the range of $I$ in Eq. \eqref{eqn:I-range} or the range of $\Theta$ in Eq. \eqref{eqn:Theta-range}, which describes the necessary condition for the occurrence of the libration of the argument of the pericenter of the perturbed object, actually depends on the ratio of the semimajor axis of the perturbed and the perturbing objects, $\alpha=a/a'$.
When deriving Eq. \eqref{eqn:cos2I-range} and Eq. \eqref{eqn:I-range}, we assumed $a \ll a'$, therefore $\alpha \to 0$.
This is equivalent to considering only the lowest order term ($2n=2$) of the averaged disturbing function \eqref{eqn:Rd-inner-j-averaged2-byu}.
In other words, we limited ourselves to use $W_{\Oalsqr}$ in Eq. \eqref{eqn:R2-final} for the disturbing Hamiltonian.
When $\alpha$ is not negligibly small, 
% we either use a higher-order expansion of the disturbing function, give up the analytical expansion, or resort to numerical quadrature (see Section 2.5).
we either use higher-order expansion of the disturbing function, or give up the analytical expansion and resort to numerical quadrature (see Section \ref{ssec:quadrature-integration}).
The numerical quadrature is to perform a double-averaging integral of the direct term of the disturbing function $R$ in Eq. \eqref{eqn:def-R-conventional-ndash} over the mean anomaly of the perturbed object $(l)$ and that of the perturbing object $(l')$ as follows:
\begin{equation}
  \frac{\mu'}{4\pi^2} \int_0^{2\pi} \int_0^{2\pi} \frac{1}{\Delta} dl dl' .
  \label{eqn:K09}
\end{equation}

By numerically evaluating the definite double integral in Eq. \eqref{eqn:K09}, \citet{kozai1962b} discovered that as $\alpha$ increases,
the upper limit of $\Theta$, below which the perturbed body's argument of pericenter is allowed to librate, also increases.
In other words, when $\alpha$ is large, objects with an even lower orbital inclination than $I \sim 39^\circ.231$ can exhibit the libration of the argument of pericenter.
This circumstance is later illustrated as Fig. \ref{fig:I02color}\textsf{a} in Appendix \ref{appen:vZLK}.

Note that if we assume $I > 90^\circ$ (the retrograde motion of the object), we can derive the following conclusion from Eq. \eqref{eqn:cos2I-range}, instead of Eq. \eqref{eqn:I-range}, as the range of inclination that allows the libration of the argument of the pericenter to happen:
\begin{equation}
  I \leq \cos^{-1} \left( -\sqrt{\frac{3}{5}} \right) = 140^\circ . 768 \cdots.
  \label{eqn:I-range-retro}
\end{equation}

Later in Section \ref{ssec:app-irregsat}, the implication that Eq. \eqref{eqn:I-range-retro} turns out to be quite significant. %{$B$3$NJ8MW%A%'%C%/!Q(B

%%%%%%%%%%%%%%%%%%%%%%%%%%%%%%%%%%%%%%%%%%%%%%%%%%%%%%%%%%
\subsection{Numerical integration\label{ssec:CR3BP-examples}}

\begin{figure*}[tbhp]\centering
 \includegraphics[width=\myfigwidthF]{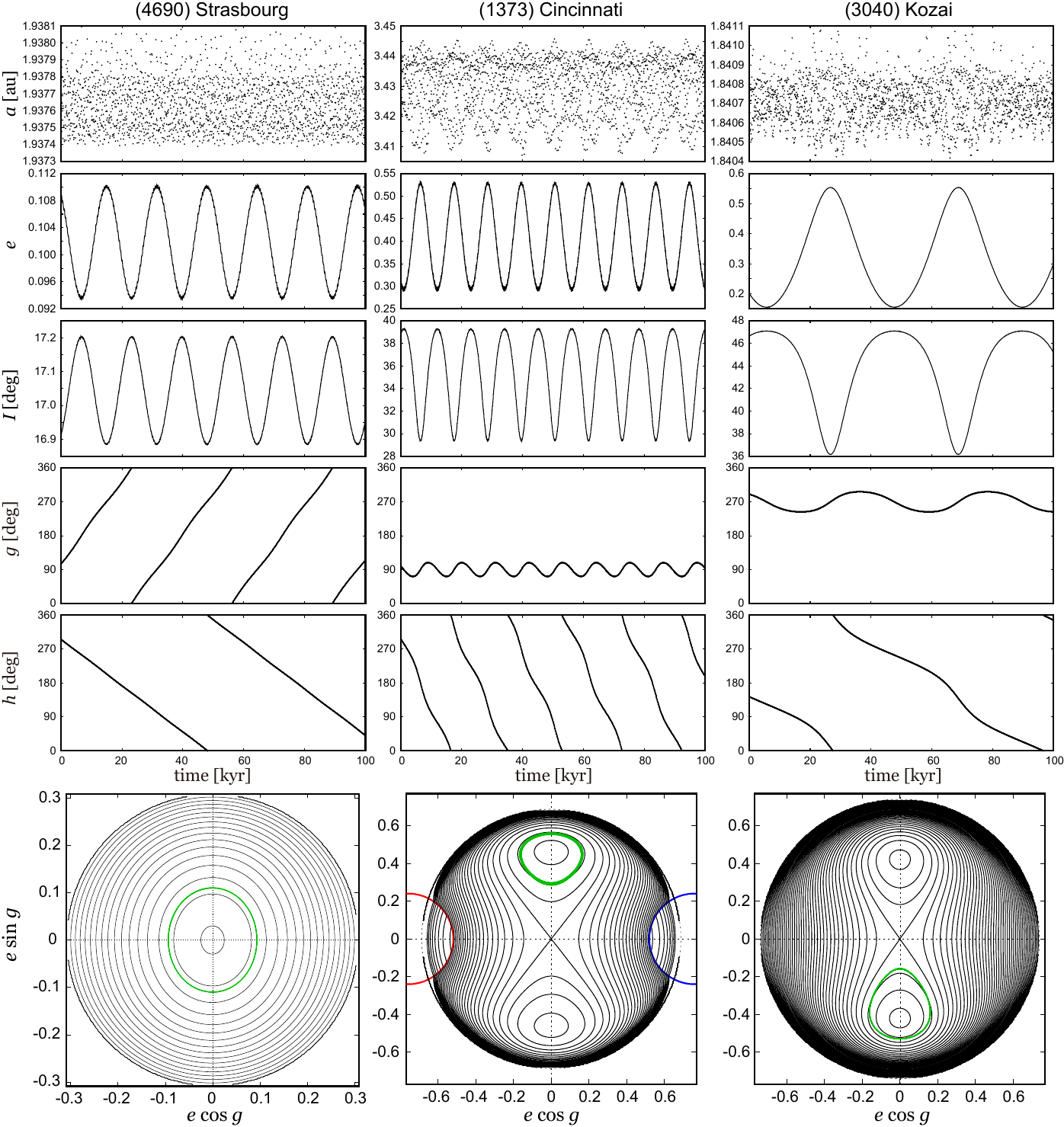} %fig2
\caption{%
(Upper five rows)
Time evolutions of the orbital elements of the three asteroid proxies.
Left:   (4690) Strasbourg.
Middle: (1373) Cincinnati.
Right:  (3040) Kozai.
(The bottom row)
The equipotential contours of the three asteroids are plotted in black on the $(e\cos g, e\sin g)$ plane of the CR3BP framework.
% The equipotential contours in black 
They are obtained by using the numerical quadrature defined in Eq. \protect\eqref{eqn:K09}.
The $(\Theta,\alpha)$ values of (4690) Strasbourg are $(\Theta, \alpha) = (0.9045, 0.3726)$.
For (1373) Cincinnati $(\Theta,\alpha) = (0.5325, 0.6569)$, and for (3040) Kozai $(\Theta,\alpha) = (0.452, 0.354)$.
The green curves denote the results of the direct numerical integration described in the upper five rows.
In other words, the numerical results plotted in the second top row for $e$ and in the fourth top row for $g$ (black curves) are converted into green curves on the $(e \cos g, e \sin g)$ plane in the bottom panels.
In the numerical integration, the initial values of the orbital elements of the asteroids' were taken from the JPL Horizons web interface as of June 7, 2017.
The initial locations of the asteroids are as follows:
$(e,g) = (0.1089756, 105^{\circ}.515364)$ for Strasbourg,
$(e,g) = (0.3151321,  99^{\circ}.948105)$ for Cincinnati, and
$(e,g) = (0.2005303, 288^{\circ}.967682)$ for Kozai.
See the main text for details of the Cincinnati panel's red and blue partial circles.
This figure was adopted from \citet[][their Figs. 2, 8, 9]{ito2019} with modification.
}
  \label{fig:CR3BP-examples}
\end{figure*}

So far, the general framework of the theory of motion of small bodies that \citet{kozai1962b} presented has been described.
We present a numerical example of the orbital evolution of three typical objects to examine the validity of this theory.
The results are presented in the five upper rows of Fig. \ref{fig:CR3BP-examples}.
Here, we consider CR3BP, in which the perturbing object (a proxy of Jupiter) has the same mass and semimajor axis as Jupiter in our solar system ($a' = 5.2042$ au).
We consider the inner problem and place three asteroid proxies as perturbed bodies orbiting inside the orbit of the Jupiter proxy.
Each of the considered asteroid proxies is intended to represent the following actual asteroids:
(4690) Strasbourg,
(1373) Cincinnati, and
(3040) Kozai.
Then, we numerically integrate their orbital evolution over 100 kyr ($10^5$ years) in the future direction by directly integrating the equations of motion \eqref{eqn:eom-relative-useR-conventional-ndash}.
The nominal step size in the numerical integration is 1 day, and the data output interval is 100 years.
For the numerical integration scheme, we employed the so-called Wisdom--Holman symplectic map \citep{wisdom1991,wisdom1992a} implemented as the \mtxtsf{SWIFT} package \citep{levison1994}.
We have added a slight improvement to this code \citep[e.g.][]{strom2005,ito2006c,ito2010}.

%%%%%%%%%%%%%%%%%%%%%%%%%%%%%%%%%%%%%%%%%%%%%%%%%%%%%%%%%%%%%%%%%%%%

Among the three columns of the panels in \mysymfigO \ref{fig:CR3BP-examples}, the motion of (4690) Strasbourg shown in the left panels exhibited the most common behavior of the objects in the inner CR3BP.
We found several noticeable characteristics in this case:
\begin{itemize}
\item Semimajor axis $(a)$ remains almost constant,
      although it exhibits a short-term oscillation with a small amplitude.
\item Eccentricity $(e)$ and inclination $(I)$ exhibit regular,
      anti-correlated oscillations.
      When $e$ becomes large (or small), $I$ becomes small (or large).
\item Argument of pericenter $(g)$ circulates in the prograde direction.
     The circulation period of this system correlates with the $e$--$I$ oscillation.
\item Longitude of ascending node $(h)$ circulates in the retrograde direction.
      Its circulation period does not appear to have any particular correlation with $e$, $I$, or $g$.
\end{itemize}
%\label{pg:CR3BP-features}

The first property ($a$ being almost constant) originates from the general fact that
the semimajor axis of the perturbed body remains constant in the doubly averaged CR3BP.
% (i.e. $\left< \left< a \right> \right>$ becomes a constant).
The second and third characteristics (the regular and correlated oscillations of $e$, $I$, and $g$) reflect the constant parameter $\Theta$.

On the other hand, the motions of the other two asteroids shown in \mysymfigO \ref{fig:CR3BP-examples} (the middle and right columns) are qualitatively different from that of (4690) Strasbourg.
Their argument of pericenter $(g)$ librates around $\frac{\pi}{2}$ or $\frac{3\pi}{2}$, instead of circulating from 0 to $2\pi$.
The oscillation still correlates with the $e$--$I$ coupling.
Regarding (3040) Kozai, the argument of pericenter $g$ seems to librate around $\frac{3\pi}{2}$ with a similar correlation.

The key to understanding the difference lies in their $\Theta$ values:
$\Theta \sim 0.91$ for (4690) Strasbourg,
$\Theta \sim 0.55$ for (1373) Cincinnati, and
$\Theta \sim 0.45$ for (3040) Kozai.
In other words, as we showed in Section \ref{ssec:Kozai-stationary}, the condition for $\Theta$ (Eq. \eqref{eqn:Theta-range}) determines whether the argument of the pericenter $g$ of each object can librate or not.

% The direct numerical integration of the equations of motion used in this paper corresponds to a procedure to numerically solve the ordinary differential equation $\DD{y}{t}=f(y;t)$ into the form of a time-dependent solution $y = y(t)$, where $t$ is time.
% Direct numerical integration is a reliable method for solving differential equations; however, it often incurs high computational costs. The numerical quadrature, described in the following subsection, is another valuable and inexpensive method.

Let us note a relevant but slightly different subject.
In the circular restricted three-body problem at the quadrupole level approximation, we can employ the following constant parameter \citep{lidov1961-en}:
\begin{equation}
  c_2 = e^2 \left( 1 - \frac{5}{2} \sin^2 I \sin^2 g \right) .
  \label{eqn:def-c2}
\end{equation}
The argument of pericenter $g$ of the perturbed object librates when $c_2 < 0$, and it circulates when $c_2 > 0$.
When $c_2 = 0$, it corresponds to a special case where the perturbed body moves along a separatrix that divides the motion of the argument of pericenter into libration and circulation in phase space.
Actually, parameters equivalent to $c_2$ in Eq. \eqref{eqn:def-c2} have been independently devised and used under different notations in later studies, such as $C_\mathrm{se}$ in \citet{kinoshita2007a}, $C_\mathrm{K}$ in \citet{katz2011}, or $C_\mathrm{KL}$ in \citet{antognini2015}.
Furthermore, \citet{vashkovyak2021} recently discussed the terminology for parameters such as $\Theta$ or $c_2$ in this line of study.
He argued that the constant parameter $c_2$ should be referred to as the Lidov integral, as it was \citet{lidov1961-en} who first derived this parameter and explicitly used it in the analysis of the doubly averaged CR3BP.

%%%%%%%%%%%%%%%%%%%%%%%%%%%%%%%%%%%%%%%%%%%%%%%%%%%%%%%%%%%%%%%%%%%%%
\subsection{Numerical quadrature\label{ssec:quadrature-integration}}
To facilitate the reader's understanding, we briefly recap the difference between the numerical quadrature that \citet{kozai1962b} dealt with and the direct numerical integration described in Section \ref{ssec:CR3BP-examples}.
The numerical quadrature is generally not directly related to the differential equations.
In general, it is an operation to find the value of a definite integral $\int_a^b f(x) dx$, where $x$ can be the time $t$ or other independent variable.
When the disturbing function is given by Eq. \eqref{eqn:def-R-conventional-ndash}, its numerical quadrature corresponds to the double averaging operation given by Eq. \eqref{eqn:K09}.
%\eqref{eqn:Delta-expansion-Legendre-inner-nge2}.
%%%
%%% $B0J2<$O(B 2.4 $B@a(B (p.24) $B$+$i$=$C$/$j0\@_$7$?$b$N$G$9!#(B
%%%
On the other hand, the direct numerical integration of the equations of motion used in this paper corresponds to a procedure of  numerically solving the ordinary differential equation $\DD{y}{t}=f(y;t)$ into the form of a time-dependent solution $y = y(t)$, where $t$ is time.
Direct numerical integration is an authentic method for solving differential equations; however, it often incurs high computational costs.
In this regard, the numerical quadrature is a useful and inexpensive method.

Unlike the direct numerical integration of the equations of motion, we cannot obtain information about the time variation of motion from the numerical quadrature.
However, this operation allows us to draw motion trajectories in the phase space, such as in the bottom panels of Fig. \ref{fig:CR3BP-examples}, particularly when the system is integrable.
This allows us to obtain a global picture of the motion without directly solving the equations of motion.
Another important aspect is that the calculation cost of the numerical quadrature in this type of study is usually much lower than that of the numerical integration.
This is because the numerical quadrature does not pursue a time-dependent solution and it is applied to systems with reduced degrees of freedom.

We now compare the results obtained through the numerical quadrature (definite integral) of the disturbing function adopted by \citet{kozai1962b} with those obtained by the direct numerical integration of the equations of motion.
We perform the numerical quadrature of the disturbing function along Eq. \eqref{eqn:K09} and draw the equipotential contours in the bottom panels of Fig. 5.
Then, we draw the result of direct numerical integration over the equipotential contours.
For simplicity, one can draw equipotential contours using the analytically obtained disturbing function truncated at a finite order.
However, numerical quadrature is expected to yield more accurate results than the truncated disturbing function, mainly when $\alpha$ is large.
\citet{kozai1962b} carried out the numerical quadrature under the name of ``numerical harmonic analysis'' (see his Fig. 1 in p. 593).
To draw the equipotential contours, we use the coordinates $\left(e\cos g, e\sin g\right)$ which is a variant of the Poincar{\'e} coordinates \citep[e.g.][]{subr2005,chenciner2015}.

The trajectories obtained by direct numerical integration of the equations of motion Eq. \eqref{eqn:eom-relative-useR-conventional-ndash} are shown as green curves in each of the bottom panels of \mysymfigO \ref{fig:CR3BP-examples}.
The equipotential contours derived by the quadrature (black) are similar to the trajectories obtained from the direct numerical integration of the equations of motion (green).
The equipotential contours of Cincinnati (bottom middle panel) possess stationary points.
This is because its $\Theta$ values are smaller than the critical value $\left(\Theta = \frac{3}{5}\right)$ in the quadrupole-level approximation.
This is also the case for the asteroid Kozai (see bottom right panel).
In contrast, $\Theta$ in Strasbourg is as large as 0.9.
With this value, the equipotential contours (bottom left panel) do not possess any local extremums except for a minimum at the origin $(0,0)$.
The argument of pericenter of the asteroid $g$ circulates regularly from 0 to $2\pi$.
Kozai's theoretical prediction is well demonstrated here.

We may find slight differences between the black and green trajectories.
This discrepancy can be attributed to the approximations used in the double-averaging operation of the disturbing function, which entirely ignores mean motion resonances.
By carrying out the orbital averaging procedure, the disturbing function becomes equivalent to an axisymmetric ring potential corresponding to an approximation of the perturbing planet as a ring along its orbit.
As a result, any mean motion resonances (commensurability) between the perturbed and perturbing bodies will lose their effect.
Although none of the three-body systems shown in \mysymfigO \ref{fig:CR3BP-examples} is trapped in a strong mean motion resonance, they may be involved with weak and high-order resonances, which the double-averaging operation neglects.

The red and blue partial circles in the bottom panel for Cincinnati in \mysymfigO \ref{fig:CR3BP-examples} represent the conditions under which the orbits of the perturbed and perturbing bodies intersect each other at the ascending node (red) and descending node (blue) of a perturbed body \citep[e.g.][]{babadzhanov1992,farinella2001,jopek2017}.
In these curves, the disturbing function becomes non-analytic and forms a set of borders on the $\left(e \cos g, e \sin g\right)$ plane (note that a non-analytic function is a function that is not locally given by a convergent power series).
The curves of orbit intersection for the other two asteroids are outside the panel ranges and are not shown in the panels for these asteroids.
The black dashed circles represent the theoretically largest eccentricity $(e_\mathrm{max})$ of the perturbed body in each CR3BP system: $e_\mathrm{max} = \sqrt{1-\Theta}$ which is realized when $\cos I = 1$.

\subsection{About the asteroid (3040) Kozai\label{ssec:asteroid-3040Kozai}}
We note that Yoshihide Kozai's prediction that the asteroid Cincinnati's $g$ librates was later confirmed by a direct numerical integration of equations of motion, including perturbations from four giant planets and Pluto \citep[][his p. 210, although no figure or table was given]{marsden1970}.
We also note that when Kozai published his paper in 1962, no asteroid was known to have the predicted librating behavior of orbits.
Years later, James G. Williams found that the argument of perihelion of stony asteroid 1979 BA, which is on a tilted orbit with a large eccentricity that exceeds Mars' orbit and the innermost region of the asteroid belt, is in libration.
He proposed it to be named (3040) Kozai because it falls in the category that Yoshihide Kozai predicted to have the librating behavior of the orbit.
See MPC (Minor Planet Circulars) 9770 (1985 July 2) or \citet{milani1989b} for more details.

%%%%%%%%%%%%%%%%%%%%%%%%%%%%%%%%%%%%%%%%%%%%%%%%%%%%%%%%%%%%%%%%%%%%%

\section{Kozai oscillation in various fields of astronomy\label{sec:applications}}
Although the ``Kozai oscillation'' was discovered in \citet{kozai1962b}, it has remained of interest among a somewhat limited community of celestial mechanics for nearly three decades.
However, with the rapid progress of observational astronomy, it gradually became evident that the "Kozai oscillation" is essential in understanding the dynamical evolution of various astronomical systems. 
In the following subsections, we discuss some examples.
Regarding the statistics of the various terms used to represent this phenomenon and our own opinion on selecting the relevant term, we prepared the Appendix \ref{appen:mainword}.

%%%%%%%%%%%%%%%%%%%%%%%%%%%%%%%%%%%%%%%%%%%%%%%%%%%%%%%%%%%%%%%%%%%%%
\subsection{Irregular satellites of giant planets\label{ssec:app-irregsat}} 
Giant planets in the solar system have many satellites.
The Hill radius $r_\mathrm{H}$ of a giant planet with mass $m_\mathrm{p}$ orbiting at a distance $a_\mathrm{p}$ from the Sun $M_\mathrm{sun}$, where the gravity of the Sun and the planet rivals, is defined as $r_\mathrm{H} = a_\mathrm{p} \left( \frac{m_\mathrm{p}}{3M_\mathrm{sun}} \right)^\frac{1}{3}$.
% \begin{equation}
%   r_\mathrm{H} = a_\mathrm{p} \left( \frac{m_\mathrm{p}}{3M_\mathrm{sun}} \right)^\frac{1}{3} .
%   \label{eqn:def-rH}
% \end{equation}
Suppose $r$ is the planetocentric distance of a satellite from its mother planet.
The satellites near the giant planet $(r \lesssim 0.05 \: r_\mathrm{H})$ are generally orbiting in circular orbits in the equatorial plane of the giant planet in the same direction (prograde) as the planet's spin.
These regular satellites are considered to have formed in the protoplanetary disk, which has been fragmented and absorbed into planetesimals and planets.
There are many additional irregular satellites in the outer region of the giant planet, $r \gtrsim 0.05 \: r_\mathrm{H}$, revolving in orbits with high eccentricity and inclinations, including some revolving in retrograde orbits.
A typical theoretical work is \citet{nesvorny2003b}, which studied the evolution of 60,000 test satellite orbits by considering the Kozai oscillation into account.
They concluded that the satellite orbits highly inclined to the ecliptic are unstable due to the Kozai oscillation.
This is because it radially stretches the satellite orbits and makes them either escape from the planetary Hill sphere, collide with massive inner moons, or impact the parent planet.

As of August 5, 2024, Jupiter is known to have 64 irregular satellites among the 72 confirmed satellites%
\footnote{\url{https://www.nao.ac.jp/new-info/satellite.html}}.
These are probably minor bodies captured by giant planets.
When the orbits of irregular satellites affected by the Kozai oscillation become highly eccentric, their pericenter may approach the planet, leading to potential collisions.
Alternatively, their apocenter can be far enough to experience significant gravitational perturbations from other planets, causing them to deviate from their CR3BP orbits.
There are 23 irregular satellites with an inclination angle of $40^{\circ} < I < 140^{\circ}$ that are probably unstable and likely to disappear in the long term due to the Kozai oscillation.

%%%
%%% Added by Takashi Ito 20241110
%%%
%%%%%%%%%%%%%%%%%%%%%%%%%%%%%%%%%%%%%%%%%%%%%%%%%%%%%%%%%%%%%%%%%%%%%
\subsection{Galactic tidal force and the Oort Cloud\label{ssec:gtforce_oort}}
There is supposed to be a reservoir of comets as a spherical shell, extending to the solar system's farthest reaches at distances between approximately 1,000 and 100,000 au, known as the Oort Cloud \citep{oort1950}.
Because the galactic plane is inclined at about $\sim 60^\circ$ to the ecliptic plane, the galactic tidal force influences the comet's motion similarly to Jupiter's gravitational influence that affects the motion of asteroids.
It can be treated as a perturbation of the Keplerian motion of the cometary system \citep[e.g.][]{heisler1986,binney1987}.
The formula of the vertical component of the galactic tidal force (relative to the galactic plane) acting on cometary motion is fundamentally equivalent to what is derived from \citeauthor{kozai1962b}'s \citeyearpar{kozai1962b} theory about the perturbation that asteroids receive from Jupiter \citep[e.g.][]{higuchi2020}.
% that of the perturbatio the asteroidal motion under Jupiter's perturbation \citep[e.g.][]{higuchi2020}.
The galactic tidal force has played a crucial role in making the outer part of the Oort Cloud dynamically isotropic, aided by random close encounters with nearby stars.
The tidal influence effectively regulates the flux and the orbital distribution of the incoming comets on a time scale of $10^9$ years \citep[e.g.][]{dones2015,ito2024}.

%%%%%%%%%%%%%%%%%%%%%%%%%%%%%%%%%%%%%%%%%%%%%%%%%%%%%%%%%%%%%%%%%%%%%
\subsection{Shrinking binary orbits\label{ssec:app-binary}}
The size of the main sequence stars shrank upon their birth from protostars that were much larger.  The presence of main sequence stars with companion stars orbiting only a few stellar radii away and recent discoveries of hot Jupiters imply that their binary orbits have shrunk by at least an order of magnitude after their formation. \citet{Fabrycky2007} studied stellar binary orbit shrinkage by the Kozai oscillation due to a distant third companion on binary orbits.
They demonstrated by direct numerical integration that binaries with periods as short as 0.1--10 days can be produced on a time scale of $10^9$ years from binaries with much more extended periods of 10 to $10^{5}$ days by the combined effect of the Kozai oscillation and tidal friction. This result is consistent with most short-period binaries having distant third companions. They also argued that this process explains the presence of a significant number of hot Jupiter population. 

%%%%%%%%%%%%%%%%%%%%%%%%%%%%%%%%%%%%%%%%%%%%%%%%%%%%%%%%%%%%%%%%%%%%%
\subsection{Type Ia supernovae\label{ssec:app-suppernova}}
Type Ia supernovae are explosions caused by merging events of two white dwarfs in close binaries that result in merged mass exceeding the Chandrasekhar limit for a stable white dwarf.
However, shrinking distant binary orbits close enough to merge within Hubble time has been an enigma.  This is because such shrinking is feasible only for a few binaries with $P < 0.3$ days.
\citet{thompson2011} found that the Kozai oscillation in triple stars helps shrink white dwarf binaries with an orbital period $P < 300$ days within Hubble time if the third star of the solar mass is in a high-inclination prograde orbit.
This effect is even more efficient for systems with a retrograde tertiary star.

%%%%%%%%%%%%%%%%%%%%%%%%%%%%%%%%%%%%%%%%%%%%%%%%%%%%%%%%%%%%%%%%%%%%%
\subsection{Blue stragglers\label{ssec:app-blue straggler}}
Blue stragglers in globular clusters are a group of enigmatic young stars populating above the turnoff point in the color-magnitude diagram.
The theory of stellar evolution shows that massive stars leave the main sequence and move onto the giant branch in color-magnitude diagrams.
Among all stars born simultaneously in the globular cluster, this process occurs from the most massive stars to the least massive stars as time passes.
Therefore, the turnoff point represents the most massive stars remaining in the main sequence of the globular cluster.
The blue stragglers should have different origins from single stars.
They might be stars that recently gained mass from stellar collisions or a merging process/mass accretion of binary stars.
\citet{perets2009a} suggested that the Kozai oscillation in triple stars could be the dominant process for forming close binaries and explaining many of the characteristics of blue stragglers, including binary fraction, period-eccentricity distribution, and their location above the turnoff point of their host clusters. 

%%%%%%%%%%%%%%%%%%%%%%%%%%%%%%%%%%%%%%%%%%%%%%%%%%%%%%%%%%%%%%%%%%%%%
\subsection{Orbit obliquity distribution of exoplanets}\label{ssec:app-exoplanets}
Spectroscopic time-series observations of exoplanet transits enable the identification of the transiting path, whether occulting from the approaching side or the receding side of the rotating host star surface.
\citet{albrecht2012} argued that the orbital rotation vectors of 14 hot Jupiters and the spin vectors of their host stars are categorized in two groups: well-aligned and misaligned from the measurements of the
Rossiter--McLaughlin effect \citep[e.g.,][]{ohta2005}.
The well-aligned (low obliquity) system could result from the star-planet tidal interactions induced by the Kozai oscillation, for which the expected tidal timescale is short.
On the other hand, the high-obliquity (misaligned or retrograde) systems have long tidal timescales and retain their initial random directions.

%%%%%%%%%%%%%%%%%%%%%%%%%%%%%%%%%%%%%%%%%%%%%%%%%%%%%%%%%%%%%%%%%%%%%
\subsection{Long-term climate change on planets \label{ssec:app-climatechange}}
As we have discussed, the Kozai oscillation produces significant variations in the eccentricity and orbital inclination of perturbed objects.
%\textcolor{red}{%
Its obliquity (the angle between the object's equatorial plane and its orbital plane) also varies.
If the perturbed object is a planet, particularly if it has climatic components such as atmosphere or ice sheet, this variation can change the distribution of incoming energy flux from its host star at different seasons and latitudes, causing climatic changes on the planet's surface.
This phenomenon is well known as the so-called Milankovitch Cycles for the Earth \citep[][see \citet{watanabe2023} for the recent research status]{milankovitch1920,milankovitch1930,milankovitch1941}.
However, the influence of the Kozai oscillation on the Milankovitch Cycles on the Earth is weak because its orbit is almost circular, and its orbital inclination remains low.
This phenomenon would be particularly remarkable on planets with large orbital eccentricities, whether they are inside or outside our solar system. 
For example, Pluto is a typical manifestation of the Kozai oscillation within our solar system with significant eccentricity and obliquity variations on timescales of several million years \citep[e.g.][]{milani1989a,kinoshita1996a,ito2002a}.
As a result, variations in the incoming solar radiation flux and its geographic distribution on Pluto induce considerable changes on its surface, together with the local accumulation or erosion of volatile ice layers \citep{young2021}.
This kind of dynamical process will also have a significant impact on the climate of extrasolar planets with large eccentricities \citep[e.g.][]{dobrovolskis2009,huang2023}.
%In recent years, studies have investigated how significant variations in eccentricity and inclination (caused by the Kozai oscillation) work in response to a drastic bifurcation and associated hysteresis of the planetary surface environment, such as those that produced the snowball status on Earth, have been addressed as a dynamical system problem \citep[e.g.][]{ji2024}.
In recent years, studies have shown that significant variations in eccentricity and inclination (caused by the Kozai oscillation) could lead to the occurrence of a drastic bifurcation and associated hysteresis of the planetary surface environment.
This type of phenomenon could have caused the snowball status on Earth \citep[e.g.][]{ji2024}.

%%%%%%%%%%%%%%%%%%%%%%%%%%%%%%%%%%%%%%%%%%%%%%%%%%%%%%%%%%%%%%%%%%%%%
\subsection{Planet around white dwarf\label{ssec:app-whitedwarf}}
White dwarfs are the degenerate cores of low-mass stars that remain after the death of these stars. It was previously thought that the survival of nearby planets was unlikely. \citet{stephan2021} pointed out that observations of accreting material from an ice giant planet onto a white dwarf (WD) and a Jupiter-sized planet transiting another WD both indicate that these planets are in close orbits well inside the stellar envelope radius of the host stars during their red giant phase. This implies that the planets must have migrated to their current orbits after their host stars had become WDs.
Furthermore, the former WD should be hot and have a short cooling time, which indicates a fast migration mechanism.
They demonstrated that the eccentric Kozai oscillation (see Section \ref{ssec:eccentric-vZLK}), combined with the stellar evolution onto WD and tidal effects, can produce the observed orbital configurations, assuming the presence of distant third stellar companions. 

%%%%%%%%%%%%%%%%%%%%%%%%%%%%%%%%%%%%%%%%%%%%%%%%%%%%%%%%%%%%%%%%%%%%%
\subsection{Black hole binaries {\textemdash} Gravitational wave\label{ssec:appl-blackhole}}
Several galaxies are expected to have a central massive black hole (MBH), numerous stellar black holes (BH), and BH-BH binaries in their centers.
These binaries form a hierarchical triple system with a central MBH, and gravitational perturbations from the MBH can cause high-eccentricity excitation in the BH-BH binary orbit.
The pericenter distance may become sufficiently small during this process, so gravitational wave emission drives the BH-BH binary to merge.
\citet{hoang2018} made simulations considering the Kozai oscillation in BH-BH binaries and demonstrated that BH-BH merging can occur. 
\citet{antonini2017} considered the formation of BH mergers through the evolution of triple stars in densely populated fields. Simulations that follow self-consistently the evolution of massive triple stars with the Kozai oscillations produced close pericenter passage binaries that could merge with gravitational-wave emission. 
  
The gravitational waves from the merging black hole binaries observed by LIGO/VIRGO showed highly asymmetric masses and significant spin that can be explained if one or both binary components are remnants of previous mergers \citep{liu2021}. 
The binaries that survived the SNe explosions generally have orbital separations too wide to merge by themselves but could merge with the aid of an external third companion, which gives rise to the Kozai oscillation.

%%%
%%% $B0KF#9';N$K$h$k8!>Z$O$3$3$^$G(B 1$B7n(B27$BF|(B($B7n(B) 12:17
%%%

%%%%%%%%%%%%%%%%%%%%%%%%%
\section{Yoshihide Kozai\label{sec:YoshihideKozai}}

%%%%%%%%%%%%%%%%%%%%%%%%%%%%%%%%%%%%%%%%%%%%%%%%%%%%%%%%%%
\subsection{As an astronomer}
Yoshihide Kozai (1928--2018) is a Japanese astronomer famous for his various works on the dynamics of small solar system bodies, planetary satellites, and artificial satellites around the Earth.
Several oral history publications are available for \citeauthor{kozai2016}'s research and life, such as \citet[][an online publication by American Institute of Physics]{devorkin1997}. 
More recently, \citeauthor{kozai2016} summarized his academic career and personal life \citep{kozai2016} including his work on the present subject.

Kozai studied celestial mechanics under the guidance of Yusuke Hagihara (1897--1979), a member of the Japan Academy, who published nine-volume textbooks on celestial mechanics \citep{hagihara1970,hagihara1972-1,hagihara1972-2,hagihara1974-1,hagihara1974-2,hagihara1975-1,hagihara1975-2,hagihara1976-1,hagihara1976-2}.
Kozai educated himself by reading classic textbooks on celestial mechanics \citep[e.g.,][]{tisserand1889,tisserand1891,tisserand1894,tisserand1896,poincare1892}.

\begin{figure}[!htbp]
\begin{center}
\includegraphics[width=\myfigwidthM]{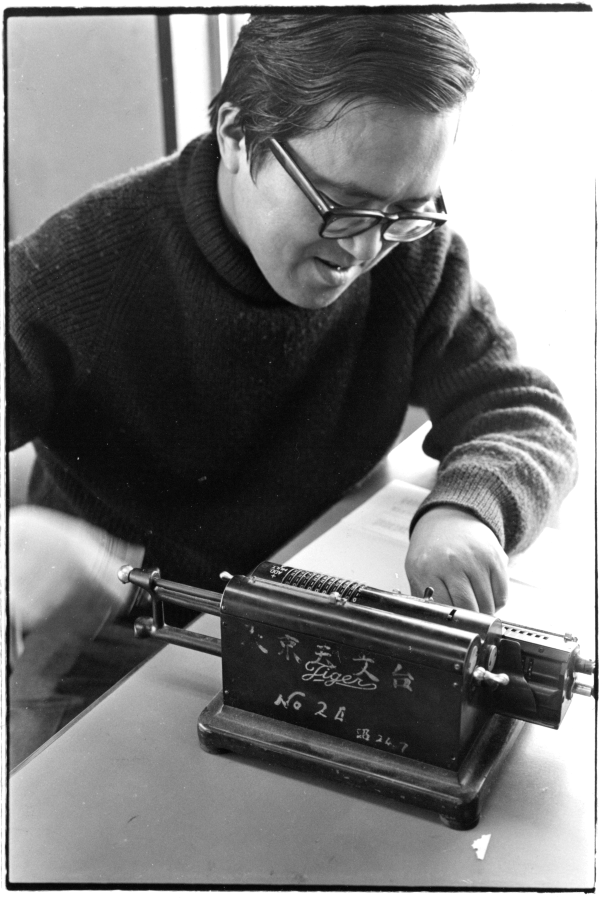}
\end{center}
\caption{%
Yoshihide Kozai demonstrates how to use a hand computer (Tiger calculator).
The Japanese characters at the back of the instrument state that this is a property of the Tokyo Astronomical Observatory.
This photo was taken in around 1955 and later given to Fumi Yoshida when he published \citet{kozai2016}. Yoshida then worked at the National Astronomical Observatory of Japan. 
This photo was transcribed from \citet[][his Fig. 1]{kozai2016}  under the permission of the publisher (Credit: Research in Astronomy and Astrophysics).}
\label{fig:Tiger}
\end{figure}

\citet{kozai1954} discussed the secular perturbation of asteroids by analyzing the effect of higher-order terms in the disturbing function, which marked his first paper on the subject.
He wrote several papers on the orbits of Saturn's inner satellites and obtained a doctorate from the University of Tokyo in 1958.
Fig. \ref{fig:Tiger} is a photo of Yoshihide Kozai using a hand computer (Tiger calculator) that he extensively used for his thesis work.
He took pride in his exceptional skill to turn the handle rapidly.

The International Geophysical Organization has proposed a campaign called ``International Geophysical Year (IGY)'' to organize international collaboration to measure the Earth from July 1957 through December 1958.
It was also when the USSR and the USA competed to develop rocket--launching technology for defense.
The USSR launched Sputnik's first artificial satellite to orbit Earth in October 1957, followed by Explorer 1 from the USA in January 1958.
Both satellites carried detectors that were useful for the IGY campaign.
The rapid development of the IGY campaign created an urgent demand for theoretical researchers qualified to conduct orbital analyses of artificial satellites. In the United States, the director of the Smithsonian Astrophysical Observatory (SAO), Fred Whipple, led a project to track the orbits of artificial satellites by deploying 12 wide-field Baker-Nunn satellite tracking cameras worldwide.
One of them was installed at the Tokyo Astronomical Observatory and Kozai worked on this campaign.
Whipple recognized the need to recruit specialists in satellite orbit studies and invited Yoshihide Kozai, who was relatively young then, to the SAO. Kozai joined the SAO satellite tracking group in October 1958.

Kozai wrote several influential articles during his four-year stay in the United States.
In \citet{kozai1959b}, he proposed formulas for the perturbed satellite orbits due to non-spherical components of the Earth's gravitational potential.
He analyzed the measured motion of satellite 1958 $\beta 2$, the Vanguard satellite. He found that the third-order spherical harmonics term of the Earth's gravitational potential, corresponding to a north-south asymmetric ``pear-shaped'' geopotential term, can reproduce well the observed variations of orbital elements with the 80-days period \citep{kozai1959b,kozai1961}.

In \citet{kozai1959b}, he proposed formulas to calculate the gravitational perturbations of the Moon and sun to the artificial satellites orbiting the Earth.
He used these formulas to predict that the lifetime of a satellite with a large eccentricity, 1959 Delta 2, for which NASA expected a lifetime as long as 20 years, would be as short as two years as its perigee height is quickly reduced due to solar perturbation \citep{kozai1959d}.
This satellite re-entered the Earth's atmosphere after a gradual decrease in its orbital altitude about 1.96 years after launch, almost exactly as \citet{kozai1959d} predicted. %
\footnote{
This satellite is better known as Explore VI (Explore 6) or the ``Paddle wheel'' as \citet{kozai1959d} mentioned in the Appendix.
However, the official designation of this satellite is 1959 Delta 1, not Delta 2 (e.g., \url{https://nssdc.gsfc.nasa.gov/nmc/spacecraft/display.action?id=1959-004A}). 
Kozai also discussed the solar radiation pressure that works to drag down artificial satellites \citep{kozai1963}.}

The technology for tracking satellite orbits evolved from optical Baker-Nunn cameras in the 1950s, radio Doppler observations in the 1960s, and laser ranging in the 1970s.
The satellite orbit calculation started with celestial mechanics formulas developed by Kozai and others, and was refined along with the development of computer technology during the following decades.
Nevertheless, the basic principles of understanding satellite orbits remain similar to those of the classic works of celestial mechanics scientists including Kozai.

%%%%%%%%%%%%%%%%%%%%%%%%%%%%%%%%%%%%%%%%%%%%%%%%%%%%%%%%%%
\subsection{As a community leader}
In addition to his scientific achievements, Kozai showed his eminent leadership in the astronomical community.
He served as the director of the Tokyo Astronomical Observatory (TAO) of the University of Tokyo (1981--1988) and modernized its operation to be open to the academic community.
In addition, he led the reorganization of TAO into an inter-university national institute.
He founded the National Astronomical Observatory (NAO) and served as its first director general (1988--1994), enabling the construction of the 8.2m Subaru Telescope on Mauna Kea.
The wide-field cameras of the Subaru Telescope, Suprime-Cam \citep{miyazaki2002} and Hyper Suprime-Cam \citep{miyazaki2018}, allowed us to find many transnepturian objects that were relevant samples to verify the dynamics described in this paper \citep[e.g.][]{ashton2023,yoshida2024,fraser2024}.
These cameras were also successful in finding the most distant galaxies to study the early history of the universe \citep[e.g.][]{iye2011}.
The Subaru Telescope revolutionally enhanced the presence of Japan's astronomical community \citep{iye2021}.
Kozai also helped establish the gravitational-wave detection group in Japan. 

He served as the president of the International Astronomical Union (IAU) from 1988 to 1991.
He was awarded the Japan Academy Prize and the Imperial Prize in 1979 for his work on celestial mechanics.
He has been a lifelong member of the Japan Academy since 1980, and served as its vice president from 2015 to 2019.
The American Astronomical Society awarded him the Brouwer Award in 1989.
Other honors include the Order of the Sacred Treasure, Gold and Silver Star (2002), and the Person of Cultural Merit (2009). 

%%%%%%%%%%%%%%%%%%%%%%%%%%%%%%%%%%%%%%%%%%%%%%%%%%%%%%%%%%%%%%%%%%%%%
\section{Discussion\label{sec:summary}}
The study of celestial mechanics has a long history.
In particular, many prominent scientists have worked on the three-body problem for hundreds of years, producing various achievements.
Among them, Kozai's work occupies an epoch-making status.
However, other scientists have previously worked on similar subjects.
The historical background of Kozai's work is summarized in Appendix \ref{appen:vZLK}.
Although the three-body problem is a fundamental and classic area of research, many issues remain unsolved.
Moreover, because it is fundamental, its development could have broad implications for mechanics and astronomy.
% As for the anti-correlated  $e$--$I$ oscillation discussed in this paper, at least two theoretical challenges remain to be addressed.
% One is the effect of the eccentricity of the perturber (i.e., when $e' > 0$).
% The second issue is the influence of the mean motion resonance between the perturbed and perturbing bodies.
As for the anti-correlated $e$--$I$ oscillation discussed in this paper, there are at least two remaining theoretical challenges to be addressed.
One is the effect of eccentricity of the perturber (i.e. when $e' > 0$).
The other is the influence of the mean motion resonance between the perturbed and perturbing bodies.

%%%%%%%%%%%%%%%%%%%%%%%%%%%%%%%%%%%%%%%%%%%%%%%%%%%%%%%%%%%%%%%%%%%%%
\subsection{Effect of perturbing body's eccentricity\label{ssec:eccentric-vZLK}}
Unlike CR3BP where the degrees of freedom can be reduced to unity through the averaging procedure, it is not easy to reduce the degrees of freedom of the system when the eccentricity of the perturbing object is non-zero $(e'>0)$.
This is because the non-zero eccentricity of the perturber makes the disturbing potential nonaxisymmetric, and the vertical component of the perturbed body's angular momentum $\left( \propto \sqrt{1-e^2} \cos I \right)$ would not remain constant anymore even in the doubly averaged system.
More specifically, the perturber's argument of pericenter $g'$ can be regarded as $g' = \pi - h$ in the disturbing function of the restricted three-body problem when $e' > 0$ \citep[][their Appendix A on their p. 7]{lithwick2011}.
Therefore, the perturbed body's longitude of ascending node $h$ remains in the disturbing function even after the quantity $h - h'$ is eliminated through Jacobi's elimination of the nodes.
Hence the conjugate momentum of $h$, $H = \sqrt{\mu a \left( 1-e^2 \right)} \cos I$, could not become constant even after the double averaging operation.
As a result, the perturbed body's dynamical behavior can be significantly different when $e'>0$.
The phenomenon is now called the eccentric \citeauthor{kozai1962b} {\mainword}.
It provides answers to many questions in the solar system dynamics that the classical framework which assumes $e'=0$ could not answer, such as the origin of objects on retrograde orbits \citep[e.g.,][]{ford2000,katz2011,naoz2016}.
The eccentric \citeauthor{kozai1962b} {\mainword} in the outer problem is also formulated \citep[e.g.,][]{naoz2017,vinson2018,deElia2019}.

Here, we highlight another issue.
Even when the orbit of the perturbing body is eccentric, its argument of pericenter $g'$ does not appear in the disturbing function if we truncate the disturbing function to the quadrupole order $(\alpha^2)$.
In this case, the system remains integrable.
For more detailed discussions, refer to \citet[][their Section 3]{naoz2013a} or \citet[][his Section 4.4.5]{shevchenko2017}.
Interestingly, \citet{lidov1976} referred to this situation as ``a happy coincidence''.
However, this coincidence no longer holds if we include the octupole-order $(\alpha^3)$ or higher terms in the disturbing function \citep[e.g.][]{farago2010,lithwick2011}.
It is worth noting the statement made by \citet[][their footnote 2 on p. 2]{lithwick2011} regarding this happy coincidence:
``However, it is perhaps more of an unhappy coincidence because it has misled some researchers into neglecting the role of the [perturbing] planet's eccentricity.''

%%%%%%%%%%%%%%%%%%%%%%%%%%%%%%%%%%%%%%%%%%%%%%%%%%%%%%%%%%%%%%%%%%%%%
\subsection{Interaction with mean motion resonance}
Mean motion resonance in astronomy refers to a circumstance where two objects' mean motion ($n$ and $n'$ in the notation in this paper) is in an integer ratio, such as 1:2 or 5:3.
These two objects are described as commensurable.
This condition makes the configuration of their orbital motions repeat relative to each other after a certain number of orbits.
Mean motion resonances, particularly lower-order ones, profoundly influence the dynamics of objects in the solar and extrasolar systems \citep[e.g.][]{malhotra2012}.

In the present paper, we assume that no mean motion resonance is working between the orbital motions of the perturbed and perturbing bodies.
However, the mean motion resonances, ubiquitous in the solar system and many other dynamical systems, are a phenomenon that continues to intrigue and engage astronomers and astrophysicists.
It is known that the Kozai oscillation can coexist with mean motion resonance, and they exhibit synergy to create more interesting dynamical phenomena \citep[e.g.][]{saillenfest2016,saillenfest2017a,saillenfest2017b}.
A typical example is Pluto's motion, which is in the 3:2 mean motion resonance with Neptune.
There is a large body of literature on the long-term orbital motion of Pluto \citep[e.g.][]{cohen1965,williams1971,nacozy1978a,sussman1988}, but we cannot fully understand its characteristics.
The Kozai oscillation in the outer problem is at work in this system.
Still, Pluto's inclination $(\sim 17^\circ)$ is not high enough to trigger the oscillation in the standard theory we introduced in the present paper \citep{malhotra2022}.
The motion of Plutinos (a group of trans-Neptunian objects, including Pluto, in the 3:2 mean motion resonance with Neptune) exhibits even greater variety.
The Kozai oscillation is also involved in many other mean motion resonances working at the outer part of the solar system and in various exoplanetary systems.
Here, we uncover a wealth of potential for theoretical research in this area, inspiring and motivating future studies.

%%%%%%%%%%%%%%%%%%%%%%%%%%%%%%%%%%%%%%%%%%%%%%%%%%%%%%%%%%%%%%%%%%%%%

\paragraph{Acknowledgments}
We thank the two anonymous reviewers for their detailed and critical feedback.
Their suggestions significantly improved the presentation and quality of the article.
This research was supported by a JSPS Grant (22K03679).
The numerical calculations used to create most figures were performed at the Center for Computational Astrophysics (CfCA), National Astronomical Observatory of Japan.
We employ the GNU Scientific Library (GSL) for the numerical quadrature presented in this paper, constituting a collection of routines for high-precision numerical computing.
The authors acknowledge the FreeBSD Project for providing a reliable, open-source operating system that facilitated the computational tasks and data analyses required for this study.
The authors used Overleaf to provide a collaborative and efficient online {\LaTeX} environment, facilitating the manuscript's preparation and formatting.
This study used NASA's Astrophysics Data System (ADS) Bibliographic Services.

\clearpage

\appendix

\section{Averaging the odd-order terms in $R$\label{appen:Pnodd}}

In Section \ref{ssec:CR3BP-R}, the expansion of the disturbing function in Eq. \eqref{eqn:Delta-expansion-Legendre-iCR3BP} contains only even terms, $2j$ ($j=1, 2, \ldots$).
We omitted the odd terms, $(2j-1)$, in the expansion of Eq. \eqref{eqn:Delta-expansion-Legendre-inner}.
This omission is justified because, in CR3BP where $e'=0$ (or $r'=a'$), all odd terms of the direct part of the disturbing function $R$ (i.e. $1/\Delta$ in Eq. \eqref{eqn:Delta-expansion-Legendre-inner-nge2}) vanish after the averaging procedure.
In this section, we demonstrate how these terms vanish and why our method is justified.
% the proposed method is justified.

We start from a generalized expression of $R$ in Eq. \eqref{eqn:Delta-expansion-Legendre-inner} where the disturbing function retains even and odd terms as follows.
We do not assume $e'=0$ yet:
\begin{equation}
\begin{aligned}
  R = \frac{\mu'}{r'}
    \sum_{j=1}^{\infty} \left( \frac{r}{r'} \right)^{j} P_{j} (\cos S) .
\end{aligned}
  \label{eqn:Delta-expansion-Legendre-iCR3BP-general}
\end{equation}

As before, we omitted the term $j=0$ from the summation in Eq. \eqref{eqn:Delta-expansion-Legendre-iCR3BP-general} because it does not depend on the perturbed body's radial distance $(r)$.
We average $R$ in Eq. \eqref{eqn:Delta-expansion-Legendre-iCR3BP-general} over the perturbing body's mean anomaly $l'$ (or its mean longitude $\lambda'$).
This can be expressed as follows:
\begin{equation}
\begin{aligned}
  \left< R \right>_{l'}
&= \frac{1}{2\pi} \int^{2\pi}_0 R dl' \\
&= \frac{1}{2\pi} \int^{2\pi}_0 
   \frac{\mu'}{r'} \sum_{j=1}^{\infty}
     \left( \frac{r}{r'} \right)^{j} P_{j} (\cos S) dl'
\end{aligned}
\label{eqn:Rd-inner-j-averaged1-general}
\end{equation}

Now, let us consider CR3BP and assume $e'=0$.
This means that $r' = a'$ (a constant), and the mean anomaly $l'$ of the perturbing body is equivalent to its true anomaly $f'$ and its eccentric anomaly $u'$.
We also assume that the perturbing body's orbit is in the reference plane such that its inclination is $I'=0$.

Here, let us note the following facts:
\begin{itemize}
%%%
\item
The perturbed body's radial distance from the central mass $r$ is independent of the perturbing body's mean anomaly $l'$ or its mean longitude $\lambda'$.
%%%
\item
The Legendre polynomials of the odd order $P_{2j-1}(x)$ are odd functions of $x$.
More specifically, they are polynomials that consist of terms containing $x, x^3, x^5, \ldots, x^{2j-1}$.
Consequently, $P_{2j-1}(\cos S)$ turn out to be polynomials composed of terms containing $\cos S, \cos^3 S, \cos^5 S, \ldots, \cos^{2j-1} S$.
See \citet[][his Section 3, p.~85]{jackson1975} for the detailed properties of the Legendre equations and the Legendre polynomials.
%%%
\item
In general, $\cos^{2j-1} \theta$ can be expressed as the sum of the terms that consist of $\cos (2j-1) \theta$ ($j=1, 2, \ldots$) through the power-reduction formulae of spherical trigonometry, such as
\begin{equation}
\begin{aligned}
  \cos^3 \theta &= \frac{ 3 \cos \theta +    \cos 3 \theta}{ 4}, \\
  \cos^5 \theta &= \frac{10 \cos \theta +  5 \cos 3 \theta +   \cos 5 \theta}{16}, \\
  \cos^7 \theta &= \frac{35 \cos \theta + 21 \cos 3 \theta + 7 \cos 5 \theta + \cos 7 \theta}{64} .
\end{aligned}
\label{eqn:ex-cospower}
\end{equation}
As a result, $P_{2j-1}(\cos S)$ turn out to be functions of $\cos S, \cos 3S, \cos 5S, \ldots, \cos (2j-1) S$.
%%%
\item
In CR3BP, $\cos S$ in Eq. \eqref{eqn:Delta-expansion-Legendre-iCR3BP-general} can be expressed as follows \citep[Eq. (7) of][note that the symbol $s$ is equivalent to our $\cos S$]{kozai1962b}:
\begin{equation}
  \cos S = \cos (f+g) \cos (f'+g') + \cos I \sin (f+g) \sin (f'+g') ,
\label{eqn:cosS-kozai1962-eq07}
\end{equation}
where $I$ is the orbital inclination of the perturbed body's orbit relative to that of the perturbing body.
In CR3BP, we can regard the quantity $f'+g'$ in Eq. \eqref{eqn:cosS-kozai1962-eq07} as being equivalent to the mean longitude of the perturbed body, $\lambda'$.
Therefore, we can rewrite Eq. \eqref{eqn:cosS-kozai1962-eq07} as follows:
\begin{equation}
  \cos S = \cos (f+g) \cos \lambda' + \cos I \sin (f+g) \sin \lambda' .
\label{eqn:cosS-kozai1962-eq07-lambda}
\end{equation}
\end{itemize}

At this point, let us pick the odd terms of the averaged $R$ from Eq. \eqref{eqn:Rd-inner-j-averaged1-general} when $e'=0$.
Denoting them as $\left< R_\mathrm{odd} \right>_{\lambda'}$, we have:
\begin{equation}
\begin{aligned}
  \left< R_\mathrm{odd} \right>_{\lambda'}
&= \frac{1}{2\pi} \int^{2\pi}_0 
     \frac{\mu'}{a'} \sum_{j=1}^{\infty}
     \left( \frac{r}{a'} \right)^{2j-1} P_{2j-1} (\cos S) d\lambda' \\
&= \frac{\mu'}{2\pi a'} \int^{2\pi}_0 \left[
            \frac{r}{a'}             P_{1} (\cos S)
+    \left( \frac{r}{a'} \right)^{3} P_{3} (\cos S)
+    \left( \frac{r}{a'} \right)^{5} P_{5} (\cos S)
+    \cdots  \right] d\lambda' .
\end{aligned}
\label{eqn:Rd-inner-j-averaged1-odd}
\end{equation}

The disturbing function $R$ depends on several variables.
However, based on the definition of $\cos S$ in Eq. \eqref{eqn:cosS-kozai1962-eq07-lambda}, we can regard that $f$, $g$, $I$ are constant in the integral in the right-hand side of Eq. \eqref{eqn:Rd-inner-j-averaged1-odd}.
This is because all of $f$, $g$, $I$ pertain to the perturbed body's motion and not to that of the perturbing body.
Consequently, we can regard that $\cos S$ in Eq. \eqref{eqn:cosS-kozai1962-eq07-lambda} depends solely on $\lambda'$ within the integral in Eq. \eqref{eqn:Rd-inner-j-averaged1-odd}.
In this case, $\cos S$ has the periodicity of $2\pi$.
Evidently, $\cos 3S, \cos 5S, \ldots, \cos (2j-1)S$ are functions only of $\lambda'$ with period of $\frac{2\pi}{3}, \frac{2\pi}{5}, \ldots, \frac{2\pi}{2j-1}$ in this integral.
This means that $P_{2j-1}(\cos S)$ has a periodicity of $\frac{2\pi}{2j-1}$ ($j=1, 2, \ldots$) in the definite integral in Eq. \eqref{eqn:Rd-inner-j-averaged1-odd} because $P_{2j-1}(\cos S)$ is a function of $\cos S, \cos 3S, \cos 5S, \ldots$, $\cos (2j-1) S$.
As a result, integrating all terms on the right-hand side of Eq. \eqref{eqn:Rd-inner-j-averaged1-odd} from 0 to $2\pi$ with respect to $\lambda'$ will yield zero for any $j$.
This gives us the conclusion that
\begin{equation}
  \left< R_\mathrm{odd} \right>_{\lambda'} = 0 .
\label{eqn:Rd-inner-j-averaged1-odd-final}
\end{equation}

Eq. \eqref{eqn:Rd-inner-j-averaged1-odd-final} validates our earlier statement that all odd terms of $R$ vanish after the averaging procedure in the CR3BP where $e'=0$.
\section{Derivation of $W_{\Oalsqr}$\label{appen:Wquad}}

In this appendix, we derive the functional form of the doubly averaged disturbing function at the quadrupole level ($W_{\Oalsqr}$ in Eq. \eqref{eqn:R2-final}) used in Section \ref{ssec:Kozai-stationary}.

We first carry out the averaging operation of the disturbing function $R$ with respect to the perturbing body's mean anomaly $l'$, followed by averaging with respect to the perturbed body's mean anomaly $l$.
In CR3BP where the orbit of the perturbing body is circular $(e'=0)$ and planar $(I'=0)$, we cannot define its ascending node or pericenter.
Therefore, the averaging operation over $l'$ is equivalent to that over the mean longitude of the perturbing body, $\lambda'$.
% Because the distance of the perturbing body from the central mass $(r' = a')$ is constant, the distance of the perturbed body from the central mass $(r)$ does not depend on the orbit of the perturbing body.
Since we presume that there is no mean motion resonance acting on the considered system, the distance between the central mass and the perturbed body $(r)$ and that between the central mass and the perturbing body $(r')$ are independent.
In addition, $r'$ is assumed to be a constant; $r' = a'$.
This explains why the average of $R_{2n}$ in Eq. \eqref{eqn:Rd-inner-j} over $l'$ has the form of Eq. \eqref{eqn:Rd-inner-j-averaged1}, where the operands $\left< \right.$ and $\left. \right>$ do not work on the coefficients $\frac{\mu'}{a'} \left( \frac{r}{a'} \right)^2$ on the right-hand side of Eq. \eqref{eqn:Rd-inner-j-averaged1}.

Now we consider the specific function of the average of $\left< P_{2n} (\cos S) \right>_{l'}$ that appears on the right-hand side of Eq. \eqref{eqn:Rd-inner-j-averaged1} and is defined in Eq. \eqref{eqn:Pj-avr}.
To proceed, we must establish two key points: the specific dependence of $\cos S$ on $l'$ (or $\lambda'$) and the specific form of each $P_{2n}$ $(n=1, 2, \ldots)$.
For the function form of $\cos S$, we consult \citet[][his Eq. (7)]{kozai1962b}.
We already showed its form in Appendix \ref{appen:Pnodd} as Eq. \eqref{eqn:cosS-kozai1962-eq07} using $f'+g'$ and also as Eq. \eqref{eqn:cosS-kozai1962-eq07-lambda} using $\lambda'$.

Many textbooks describe the characteristics of the Legendre polynomials $P_{2n}(x)$ in detail \citep[e.g.][his Section 3.2, p. 85]{jackson1975}.
When $2n=2$, the function takes the following form:
\begin{equation}
  P_2 (x) = \frac{1}{2} \left( 3 x^2 - 1\right) ,
\label{eqn:P2-x}
\end{equation}
which gives
\begin{equation}
  P_2 (\cos S) = \frac{1}{2} \left( 3 \cos^2 S - 1\right) .
\label{eqn:P2-cosS}
\end{equation}

By averaging the both sides of Eq. \eqref{eqn:P2-cosS} with respect to $l'$, we obtain
\begin{equation}
\begin{aligned}
  \left< P_2 (\cos S) \right>_{l'}
& = \frac{1}{2} \left( 3 \left< \cos^2 S \right>_{l'}     - 1 \right) \\
& = \frac{3}{2} \int_0^{2\pi} \cos^2 S d l' - \frac{1}{2} .
\end{aligned}
\label{eqn:P2-avr-def}
\end{equation}

As mentioned above, we can replace $dl'$ on the right-hand side of Eq. \eqref{eqn:P2-avr-def} with $d \lambda'$ in the considered system.
Thus, we can rewrite Eq. \eqref{eqn:P2-avr-def} as follows:
\begin{equation}
\begin{aligned}
  \left< P_2 (\cos S) \right>_{\lambda'}
& = \frac{1}{2} \left( 3 \left< \cos^2 S \right>_{\lambda'}     - 1 \right) \\
& = \frac{3}{2} \int_0^{2\pi} \cos^2 S d \lambda' - \frac{1}{2} .
\end{aligned}
\label{eqn:P2-avr-def-lambda}
\end{equation}

We now calculate the specific form of $\cos^2 S$ using Eq. \eqref{eqn:cosS-kozai1962-eq07-lambda} and prepare for its averaging as follows
\begin{equation}
\begin{aligned}
  \cos^2 S
&= \left( \cos (f+g) \cos \lambda' + \cos I \sin (f+g) \sin \lambda' \right)^2 \\
&=    \cos^2 (f+g) \cos^2 \lambda' \\
& \quad
  + 2 \cos   (f+g) \sin (f+g) \cos I \cos \lambda' \sin \lambda' \\
& \quad\quad
  + \sin^2 (f+g) \cos^2 I\sin^2 \lambda'  .
\end{aligned}
\label{eqn:cosS-square-1}
\end{equation}

Using the expression of $\cos^2 S$ in Eq. \eqref{eqn:cosS-square-1} and employing the usual trigonometric formulas of $\lambda'$ such as
\begin{equation}
  \cos^2 \lambda' = \frac{1 + \cos 2 \lambda'}{2}, \quad
  \sin^2 \lambda' = \frac{1 - \cos 2 \lambda'}{2}, \quad
  \cos \lambda' \sin \lambda' = \frac{\sin 2 \lambda'}{2} ,
\label{eqn:lambda-trigonometry}
\end{equation}
we obtain the following expression for the average of $\cos^2 S$ with respect to $\lambda'$:

\begin{equation}
\begin{aligned}
  \left< \cos^2 S \right>_{\lambda'}
&= \frac{1}{2\pi} \int_0^{2\pi} \cos^2 S d \lambda \\
&= \frac{1}{2\pi} \int_0^{2\pi} \left[
   \cos^2 (f+g) \frac{1+\cos 2\lambda'}{2}
 + \cos (f+g) \sin (f+g) \cos I \sin 2 \lambda' \right. \\
& \quad \quad \left.
 + \sin^2 (f+g) \cos^2 I \frac{1-\cos 2\lambda'}{2} 
                 \right] d\lambda' \\
&= \frac{1}{2\pi} \int_0^{2\pi} \frac{1}{2} \left[
  \cos^2 (f+g) (1+\cos 2\lambda')
+ \cos I \sin 2(f+g) \sin 2 \lambda' \right. \\
& \quad \quad \left.
+ \sin^2 (f+g) \cos^2 I (1-\cos 2\lambda') \right] d\lambda' \\
&= \frac{1}{2} \left[
  \cos^2 (f+g)
+ \sin^2 (f+g) \cos^2 I
               \right] \\
&= \frac{1}{4} \sin^2 I \cos 2 (f+g) - \frac{1}{4} \sin^2 I + \frac{1}{2} .
\end{aligned}
\label{eqn:cos2S-avr}
\end{equation}

By substituting Eq. \eqref{eqn:cos2S-avr} into Eq. \eqref{eqn:P2-avr-def-lambda}, we obtain
\begin{equation}
\begin{aligned}
  \left< P_2 (\cos S) \right>_{\lambda'}
&= \frac{3}{2} \left< \cos^2 S \right>_{\lambda'} - \frac{1}{2} \\
&= \frac{3}{8} \sin ^2 I \cos 2 (f+g) - \frac{3}{8} \sin^2 I + \frac{1}{4} .
\end{aligned}
\label{eqn:P2-avr-2}
\end{equation}

We now perform the averaging operation of the disturbing function with respect to the perturbed body's mean anomaly $l$ (or by its mean longitude $\lambda$).
By setting $2n=2$ and substituting the result of Eq. \eqref{eqn:P2-avr-2} into Eq. \eqref{eqn:Rd-inner-j-averaged2-byu}, we have the following relationship:
\begin{equation}
\begin{aligned}
  \left< \left< R_{2} \right>_{\lambda'} \right>_{l}
& = \frac{\mu'}{a'} \left( \frac{a}{a'} \right)^{2}
      \frac{1}{2\pi} \int^{2\pi}_0
        \left( \frac{r}{a} \right)^{2} \left< P_{2} (\cos S) \right>_{\lambda'} dl \\
& = \frac{\mu'}{a'} \left( \frac{a}{a'} \right)^{2}
      \frac{1}{2\pi} \int^{2\pi}_0
        \left( \frac{r}{a} \right)^{2} 
          \left[
  \frac{3}{8} \sin ^2 I \cos 2 (f+g) - \frac{3}{8} \sin^2 I + \frac{1}{4} 
          \right] dl \\
& = \frac{\mu'}{a'} \left( \frac{a}{a'} \right)^{2}
      \frac{1}{2\pi} \int^{2\pi}_0
        \left( \frac{r}{a} \right)^{2} 
          \frac{3}{8} \sin ^2 I \cos 2 f \cos 2 g dl \\
& \quad
  - \frac{\mu'}{a'} \left( \frac{a}{a'} \right)^{2}
      \frac{1}{2\pi} \int^{2\pi}_0
        \left( \frac{r}{a} \right)^{2} 
          \frac{3}{8} \sin ^2 I \sin 2 f \sin 2 g dl \\
& \quad \quad
  + \frac{\mu'}{a'} \left( \frac{a}{a'} \right)^{2}
      \frac{1}{2\pi} \int^{2\pi}_0
        \left( \frac{r}{a} \right)^{2} 
          \left( - \frac{3}{8} \sin^2 I + \frac{1}{4} \right) dl .
\end{aligned}
\label{eqn:Rd-inner-j-averaged2-byu-more}
\end{equation}

In what follows, we evaluate each term on the right-hand side of Eq. \eqref{eqn:Rd-inner-j-averaged2-byu-more}.
For this purpose, we convert the integral variable from the mean anomaly $l$ to the eccentric anomaly $u$ by employing the differential relationship between $l$ and $u$ \citep[e.g.][their p.~22]{brouwer1961}
\begin{equation}
  \DD{l}{u} = \frac{r}{a} .
\label{eqn:DlDu-def}
\end{equation}
We also use the geometric relationship between the eccentric anomaly and the true anomaly given in Eq. \eqref{eqn:cosf-sinf-by-u} at Section \ref{ssec:two-body}, as well as the relationship between $r$, $a$, and $u$, which was also given in Eq. \eqref{eqn:r-by-u}.

Among the three terms on the right-hand side of Eq. \eqref{eqn:Rd-inner-j-averaged2-byu-more}, first let us pick the term that includes $\cos 2f$ as a factor.
Employing the usual trigonometric formulas such as
\begin{equation}
  \cos^3 u = \frac{3 \cos u + \cos 3 u}{4}, \quad
  \cos^2 u = \frac{1 + \cos 2 u}{2} ,
\label{eqn:u-trigonometry}
\end{equation}
the essential part of the $\cos 2f$ term can be expressed as follows:
\begin{equation}
\begin{aligned}
&     \frac{1}{2\pi} \int^{2\pi}_0
        \left( \frac{r}{a} \right)^{2} \cos 2 f dl \\
&=
      \frac{1}{2\pi} \int^{2\pi}_0
        \left( \frac{r}{a} \right)^{2} \left( 2 \cos^2 f -1 \right) \DD{l}{u} du \\
&=
      \frac{1}{2\pi} \int^{2\pi}_0
        \left( \frac{r}{a} \right)^{2} \left[ 2 \left( \frac{\cos u - e}{1 - e \cos u} \right)^2 - 1 \right] \left( \frac{r}{a} \right) du \\
&=
      \frac{1}{2\pi} \int^{2\pi}_0
        \left( 1 - e \cos u \right)^3
        \left[ 2 \left( \frac{\cos u - e}{1 - e \cos u} \right)^2 - 1 \right] du \\
&= \frac{1}{2\pi} \int^{2\pi}_0 \left[
         2 e^2 - 1
 - \left(2 e^3 + e  \right) \cos u \right. \\
& \quad\quad \left.
 + \left(  e^2 + 2  \right) \frac{1+\cos 2u}{2}
 + \left(  e^3 - 2e \right) \frac{3 \cos u + \cos 3u}{4}
  \right] du \\
&= 2 e^2 - 1 + \frac{e^2 + 2}{2} \\
&= \frac{5 e^2}{2} .
\end{aligned}
\label{eqn:ra2cos2f-avr-1}
\end{equation}

Next, we deal with the second term on the right-hand side of Eq. \eqref{eqn:Rd-inner-j-averaged2-byu-more} which includes $\sin 2f$ as a factor.
The essential part of this term results in the following expression:
\begin{equation}
\begin{aligned}
&     \frac{1}{2\pi} \int^{2\pi}_0
        \left( \frac{r}{a} \right)^{2} \sin 2 f dl \\
&=    \frac{1}{2\pi} \int^{2\pi}_0
        \left( \frac{r}{a} \right)^{2} 2 \sin f \cos f \left( \frac{r}{a} \right) du \\
&=    \frac{1}{2\pi} \int^{2\pi}_0
      2 \left( \frac{r}{a} \right)^{2}
        \left( \frac{\sqrt{1-e^2}\sin u}{1 - e \cos u} \right)
        \left( \frac{\cos u - e}{1 - e \cos u}         \right)
        \left( \frac{r}{a} \right) du \\
&=    \frac{1}{2\pi} \int^{2\pi}_0
      2 \left( \frac{r}{a} \right)^{2}
        \left( \frac{a}{r} \right)^{2}
        \sqrt{1-e^2} \sin u \left( \cos u - e \right)
        \left( \frac{r}{a} \right) du \\
&=    \frac{2\sqrt{1-e^2}}{2\pi} \int^{2\pi}_0
        \left( 1 - e \cos u \right)
        \left( \sin u \cos u - e \sin u \right) du \\
&=    \frac{2\sqrt{1-e^2}}{2\pi} \int^{2\pi}_0
 \left(
   \sin u \cos u - e \sin u - e \sin u \cos^2 u + e^2 \sin u \cos u
 \right) du \\
&=     \frac{2\sqrt{1-e^2}}{2\pi} \int^{2\pi}_0
 \left(
   -\frac{5e}{4}    \sin  u
   +\frac{1+e^2}{2} \sin 2u
   -\frac{e}{4}     \sin 3u
 \right) du \\
&= 0 .
\end{aligned}
\label{eqn:ra2sin2f-avr-1}
\end{equation}

The last term on the right-hand side of Eq. \eqref{eqn:Rd-inner-j-averaged2-byu-more} does not contain $f$ or $u$.
The essential part of this term results in the following expression:
\begin{equation}
\begin{aligned}
&     \frac{1}{2\pi} \int^{2\pi}_0
        \left( \frac{r}{a} \right)^{2} dl \\
&=    \frac{1}{2\pi} \int^{2\pi}_0
        \left( \frac{r}{a} \right)^{2} \left( \frac{r}{a} \right) du \\
&=    \frac{1}{2\pi} \int^{2\pi}_0
        \left( \frac{r}{a} \right)^{3} du \\
&=    \frac{1}{2\pi} \int^{2\pi}_0
        \left( 1 + e \cos u \right)^{3} du \\
&=    \frac{1}{2\pi} \int^{2\pi}_0
  \left( 1 + 3e \cos u + 3e^2 \cos^2 u + e^3 \cos^3 u \right) du \\
&=    \frac{1}{2\pi} \int^{2\pi}_0
  \left( 1 + 3e \cos u + 3e^2 \frac{1+\cos 2u}{2} + e^3 \frac{3\cos u + \cos 3u}{4} \right) du \\
&=    \frac{1}{2\pi} \int^{2\pi}_0 \left[
   1 + \frac{3e^2}{2}
 + \left( 3e + \frac{3e^2}{4} \right) \cos  u
 +             \frac{3e^2}{2}         \cos 2u
 +             \frac{ e^2}{4}         \cos 3u
                                   \right] du \\
&= 1 + \frac{3 e^2}{2} .
\end{aligned}
\label{eqn:ra2-avr-1}
\end{equation}

Substituting the results given in Eqs. \eqref{eqn:ra2cos2f-avr-1}, \eqref{eqn:ra2sin2f-avr-1}, and \eqref{eqn:ra2-avr-1} into the right-hand side of Eq. \eqref{eqn:Rd-inner-j-averaged2-byu-more}, we obtain the final expression of $\left< \left< R_{2} \right>_{\lambda'} \right>_{l}$ as follows:
\begin{equation}
\begin{aligned}
  \left< \left< R_{2} \right>_{\lambda'} \right>_{l}
& = \frac{\mu'}{a'} \left( \frac{a}{a'} \right)^{2}
      \frac{1}{2\pi} \left( \frac{3}{8} \sin ^2 I \cos 2 g \right)
        \int^{2\pi}_0 \left( \frac{r}{a} \right)^{2} \cos 2 f dl \\
& \quad
  - \frac{\mu'}{a'} \left( \frac{a}{a'} \right)^{2}
      \frac{1}{2\pi} \left( \frac{3}{8} \sin ^2 I \sin 2 g \right)
        \int^{2\pi}_0 \left( \frac{r}{a} \right)^{2} \sin 2 f dl \\
& \quad \quad
  + \frac{\mu'}{a'} \left( \frac{a}{a'} \right)^{2}
      \frac{1}{2\pi}  \left( - \frac{3}{8} \sin^2 I + \frac{1}{4} \right) 
        \int^{2\pi}_0 \left( \frac{r}{a} \right)^{2} dl \\
&= \frac{\mu'}{a'} \left( \frac{a}{a'} \right)^{2}
\left[
         \frac{3}{8} \sin^2 I \cos 2g \cdot \frac{5e^2}{2}
- \left( \frac{3}{8} \sin^2 I - \frac{1}{4} \right) \left( 1 + \frac{3e^2}{2} \right) 
\right] \\
&= 
\frac{1}{16}
\frac{\mu'}{a'} \left( \frac{a}{a'} \right)^2 
\left[
  15 e^2 \sin^2 I \cos 2 g - \left(3e^2+2 \right) \left( 3\sin^2 I -2 \right)
\right] .
\end{aligned}
  \label{eqn:R2-final-true}
\end{equation}

The result expressed as $\left< \left< R_{2} \right>_{\lambda'} \right>_{l}$ in Eq. \eqref{eqn:R2-final-true} is equivalent to $W_{\Oalsqr}$ in Eq. \eqref{eqn:R2-final}, differing only in the leading sign and the mass factor $(\mu' = {\cal G} m')$.

The higher-order terms of the disturbing function $R$ can be derived in a similar manner for both inner and outer problems.
\citet{ito2016} obtained the function forms of the doubly averaged disturbing function up to $2n=12$ for the inner CR3BP and up to $2n=14$ for the outer CR3BP.

\section{Choice of terms \label{appen:mainword}}

This section briefly considers the most relevant term to describe the phenomenon Kozai investigated.
\citet[][their subsection 6.2.5]{ito2019} conducted an extensive study on the choice of words describing this phenomenon in the literature. 
Their survey was among the abstracts of refereed and non-refereed publications in astronomy and physics registered in NASA's Astrophysics Data System (ADS, \url{https://ui.adsabs.harvard.edu/}). They searched for the keywords,
``\textit{Kozai ---\/}'' and ``\textit{Lidov ---\/}''
where ``---'' represents one of the following terms:
\textit{mechanism,\/}
\textit{resonance,\/}
\textit{cycle,\/}
\textit{oscillation,\/}
\textit{effect,\/}
\textit{dynamics,\/}
\textit{perturbation,\/} or
\textit{libration.\/}
Although it was a crude statistical method, they found a general tendency as to which term the authors in the field preferred to describe these secular dynamics.
Among the words searched, the most popular terms were
 \textit{mechanism,\/}
 \textit{resonance,\/}
 \textit{cycle,\/}
 \textit{oscillation,\/} and
 \textit{effect\/} listed in order of frequency.
% Terms \textit{perturbation,\/} \textit{dynamics,\/} and  \textit{libration,\/} were not popular.
In conducting the survey, \citeauthor{ito2019} recommended using \textit{{\mainword}\/}, despite \textit{mechanism,\/} being most frequently used to express the subject in past literature.
They reasoned that Kozai's work primarily focuses on the apparent phenomenon itself (\textit{oscillation\/}) rather than on a detailed investigation of its underlining (\textit{mechanism\/}).
The terms \textit{oscillation} and \textit{cycle} express a phenomenon, whereas \textit{mechanism} and \textit{effect} are generic words lacking a concrete image.
We followed their opinion and have used ``\textit{{\mainword}\/}'' in the title of this paper.

% Meanwhile the term \textit{resonance\/} has also been used commonly
On the other hand, although the term \textit{resonance\/} has also been commonly used \citep[e.g.][]{marsden1999,thompson2011,deElia2019}, \citet{ito2019} highlighted that this term is not entirely appropriate for denoting this subject.
This is because a resonance is supposed to work between a frequency associated with one object and a frequency associated with another object.
However, the primary focus is the motion of the argument of pericenter of the perturbed body, whereas no such motion can be defined for the perturbing body's circular orbit in CR3BP.
Namely, only the frequency of a single object is involved here; therefore, it is not appropriate to call this oscillation a resonance.
%Nevertheless, people have different views regarding the issue of whether this oscillation is a resonance or not:
% Some claim that it is certainly a kind of resonance, but others do not.
However, some claim that it is indeed a resonance. A reason for this is as follows.
As described, the typical outcome of this oscillation is that the argument of the pericenter of the perturbed object, $g$, undergoes libration.
This can be expressed as follows:
\begin{equation}
  \DD{g}{t} \sim 0 .
  \label{eqn:def-g-libration}
\end{equation}
This equation can be rewritten using the longitude of the ascending node $h$ and the longitude of pericenter $\varpi = g + h$ as follows:
\begin{equation}
  \DD{\varpi}{t} \sim \DD{h}{t} .
    \label{eqn:def-g-libration-2}
\end{equation}

Eq. \eqref{eqn:def-g-libration-2} indicates that the variation rates of $\varpi$ and $h$ are in the 1:1 commensurability, and one may regard that they are in a resonance.

% When $e'=0$ the argument that the system has only one frequency holds. However, in the limit of infinitesimally small value of $e'$, the argument of commensurability holds, and we are in favor of regarding this as a resonance.

\section{Works by Lidov and von Zeipel\label{appen:vZLK}}
\subsection{Relation of the three works}
As is well known now, an equivalent work to what \citet{kozai1962b} did was published slightly earlier by a Soviet Union scientist, Michail L'vovich Lidov, in 1961.
Both \citet{lidov1961-en} and \citet{kozai1962b} worked on the inner problem of CR3BP, with neither addressing the outer problem in which the perturbed body orbits outside the perturber.
\citet{lidov1961-en} dealt only with the quadrupole-order approximation, but he provided a detailed analysis of the system behavior at the border between the circulation and libration of $g$.
His analysis result is summarized as a plot which is now called the Lidov diagram \citep[e.g.][their Fig. 10 in their p. 35]{ito2019}.
This diagram is quite useful to understand the dynamical nature of the doubly averaged CR3BP.

Yoshihide Kozai and Michail Lidov met once in 1961 at a conference in Moscow \citep[e.g.][]{grebenikov1962-en}, and discussed this subject \citep[][their p. 90]{ito2019}.
\citet{kozai1962b} cites the presentation given by \citeauthor{lidov1961-en} at the conference, so it would be fair to say that \citeauthor{kozai1962b}'s work and \citeauthor{lidov1961-en}'s work are not independent.
For this reason, the $e$--$I$ oscillation that we discussed in this paper was called the Kozai--Lidov or Lidov--Kozai oscillation.

However, a recent extensive survey of literature from the late 19th to early 20th century \citep{ito2019} revealed that pioneering work on this subject had already been established during that era.
A Swedish astronomer, Edvard Hugo von Zeipel, accomplished this.
\citeauthor{vonzeipel1910}'s achievement at that time encompasses most of the fundamental ingredients Kozai and Lidov published $\sim$70 years later.
Let us briefly summarize the work by \citeauthor{vonzeipel1910}.
\begin{itemize}
\item
\citeauthor{vonzeipel1910} brought the Lindstedt series \citep{lindstedt1882a,lindstedt1882b,lindstedt1883a,lindstedt1884,poincare1886,poincare1892,bohlin1888,gylden1891,gylden1893} into canonical perturbation theory.
The Lindstedt series is also known as the Lindstedt–Poincaré method.
In general, direct application of the classical perturbation theory to a weakly perturbed system produces artificial secular terms that tend to increase indefinitely.
The Lindstedt series provides a theoretical framework for systematically suppressing the occurrence of the secular terms \citep[e.g.][]{lichtenberg1992}. 
\citeauthor{vonzeipel1910} applied the Lindstedt series to the doubly averaged CR3BP to locate the possible local extremums of the disturbing function in addition to the origin $(e \cos g, e \sin g) = (0, 0)$.
It is probably a work product of his later famous works, currently known as the \citeauthor{vonzeipel1910} method in the canonical perturbation theory \citep{vonzeipel1916a,vonzeipel1916b,vonzeipel1917a,vonzeipel1917b}.
\item
\citeauthor{vonzeipel1910} dealt both with inner and outer problems because his subject was the orbital motion of long-period comets, which can take either of the two forms.
Subsequent to \citeauthor{vonzeipel1910}, no one accomplished the detailed treatment of the outer problem until the 1990s when trans-Neptunian objects and exoplanets were discovered.
\item
\citeauthor{vonzeipel1910} considered the case of the orbit intersection between perturbed and perturbing objects. This subject was analytically considered again only at the end of the 20th century,
\item
\citeauthor{vonzeipel1910} gave general discussions on approximation higher than the quadrupole-order $(\alpha^2)$ and yielded solutions as accurate as modern studies.
% Remember that the approximation that \citep{kozai1962b} and \citep{lidov1961} used was at the $\alpha^2$ order.
\item
\citeauthor{vonzeipel1910} showed equipotential trajectories in his 1910 article. Although they appear to be handwritten, they are quite accurate.
\item
\citeauthor{vonzeipel1910} predicted the existence of stable orbits of small bodies around Neptune's orbit.
We can interpret this as a prediction of trans-Neptunian objects or Centaurs.
We should remember that even Pluto was not discovered in his lifetime.
\end{itemize}

\begin{figure}[!htbp]
\begin{center}
\includegraphics[width=\myfigwidthM]{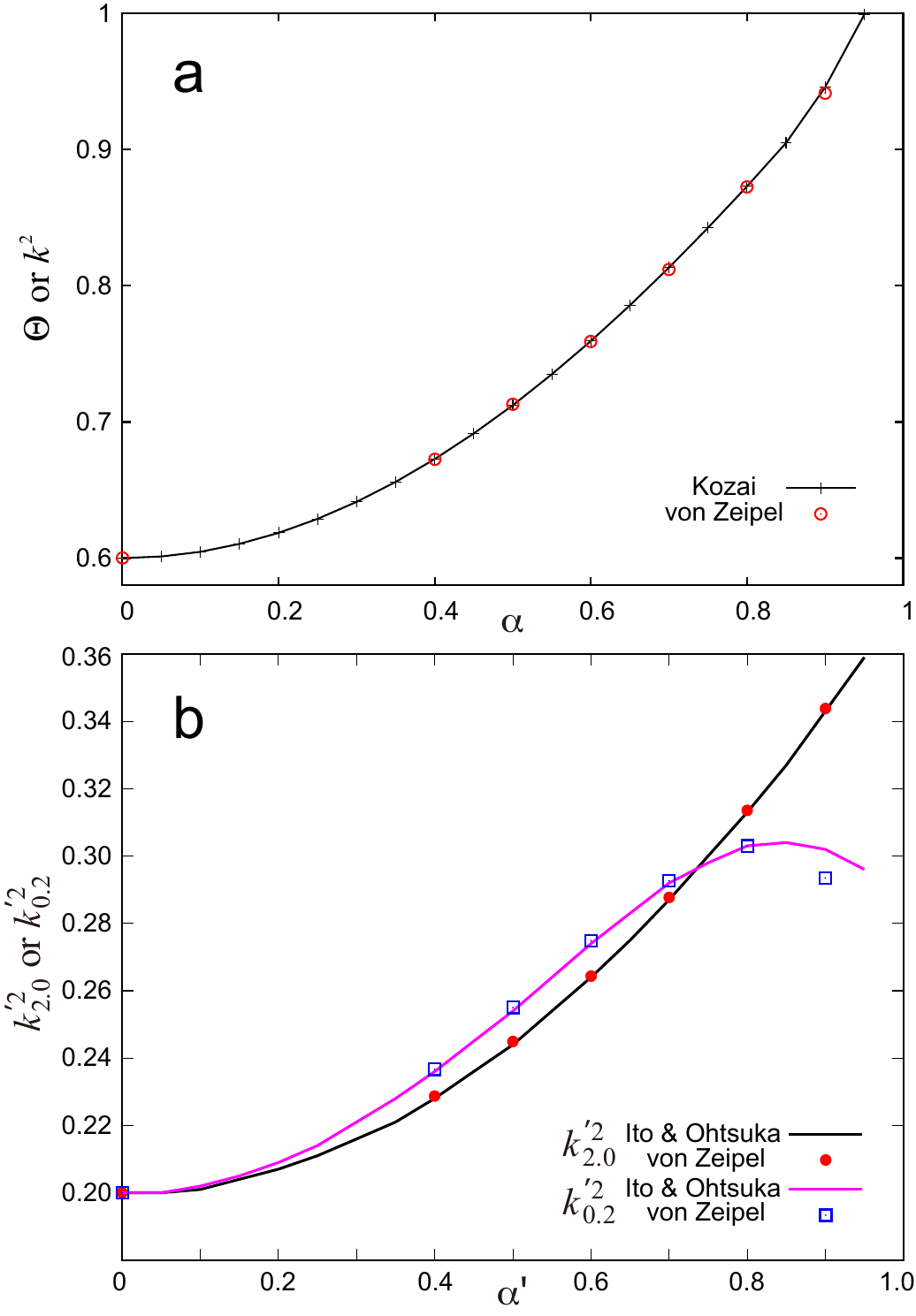}
\end{center}
\caption{%
\mtxtsf{a}:
Dependence of \citeauthor{kozai1962b}'s $\Theta$ and \citeauthor{vonzeipel1910}'s $k^2$ on $\alpha$ in the inner problem.
These values denote the threshold of the libration of the perturbed object's $g$.
At each $\alpha$, the perturbed object's $g$ can librate if the object's $\Theta$ (or $k^2$) is lower than the plotted values.
The black curves with $+$ denote the numerically calculated values of $\Theta$ in \citet{kozai1962b}.
The red open circles denote the values of $k^2$ tabulated in \citet{vonzeipel1910}.
\mtxtsf{b}: The threshold value of the libration of the perturbed object's $g$ in the outer problem represented as
${k'}^2_{2.0}$ (the black solid curve) and
${k'}^2_{0.2}$ (the magenta solid curve) along with their dependence on $\alpha'$ that \citet{ito2019} calculated.
% were calculated by \citet{ito2019}.
The tabulated results in \citet{vonzeipel1910} are converted into
${k'}^2_{2.0}$ (the filled red circle) and
${k'}^2_{0.2}$ (the open blue square with a central dot), and plotted them in this panel.
The panel \mtxtsf{b} is taken from \citet[][their Figs. 22a]{ito2019}.
}
\label{fig:I02color}
\end{figure}

Let us now quantitatively compare von Zeipel's theory with Kozai's and more recent works to examine its accuracy.
An example subject chosen is the dependence of the maximum (threshold) value of $\Theta$ that allows the libration of the argument of pericenter $g$ on the ratio of the semimajor axis, $\alpha = a/a'$.
\citet{kozai1962b} estimated this dependence in the inner problem by numerical quadrature (see the third column of \citeauthor{kozai1962b}'s Table I in his p. 592, as well as \citeauthor{kozai1962b}'s \mysymfigO 1 in his p. 593).
\citet{vonzeipel1910} used higher-order approximation and carried out similar calculations.
\citet{vonzeipel1910} used a notation $k^2$ instead of $\Theta$.
The comparison results are presented in the panel \textsf{a} of \mysymfigO \ref{fig:I02color}.
This tells us that \citeauthor{vonzeipel1910}'s $\Theta$ almost closely aligns with \citeauthor{kozai1962b}'s $k^2$ up to $\alpha = 0.9$.

We should remark here that the dependence of $\Theta$'s threshold value on $\alpha$ in the inner problem was theoretically established earlier by \citet{vonzeipel1901}.
For the threshold value $\Theta$ or $k^2$ in the limit of $\alpha \to 0$ (that is, $\Theta = \frac{3}{5}$), \citet{vonzeipel1898} had already found this value at the end of the nineteenth century.

%%%%%%%%%%%%%%%%%%%%%

A significant feature of von Zeipel's work is that he dealt with the outer problem with the same precision as the inner problem.
Not only did Kozai and Lidov not construct a theory along these lines for the outer problem, but it is even fair to say that, after von Zeipel's work, no proper theory was published until the 1990s.
From the theoretical viewpoint, the outer problem is more complex and challenging than the inner problem.
For example, in the inner problem, as we have already mentioned, the equilibrium point of the argument of the pericenter $g$ occurs only when $\cos g = -1$ (i.e., $g=\pm \frac{\pi}{2}$).
On the other hand, in the outer problem, the equilibrium point of $g$ also occurs when $\cos g = +1$ (i.e., $g=0$ or $g=\pi$).
Therefore, the thresholds that produce the libration of $g$ are also divided into two branches.
\citet{vonzeipel1910} denoted these threshold values as $k'^2_{2.0}$ (for $\cos g = +1$) and $k'^2_{0.2}$ (for $\cos g = -1$).
\mysymfigO \ref{fig:I02color}\mtxtsf{b} shows the dependence of ${k'}^2_{2.0}$ and ${k'}^2_{0.2}$ on $\alpha' (\equiv \alpha^{-1} = a'/a)$.
For comparison, we also plotted the result of a recent numerical quadrature \citep{ito2019} for the same quantities.
The panel \mtxtsf{b} indicates that the result of \citeauthor{vonzeipel1910} agrees with the modern numerical results excellently, with only a slight deviation in the ${k'}^2_{0.2}$ branch for $\alpha' > 0.80$.

%%%%%%%%%%%%%%%%%%%%%%%%%%%%%%%%%%%%%%%%%%
\subsection{Citation analysis}
\begin{figure}[!htbp]
\begin{center}
 \includegraphics[width=\myfigwidthF]{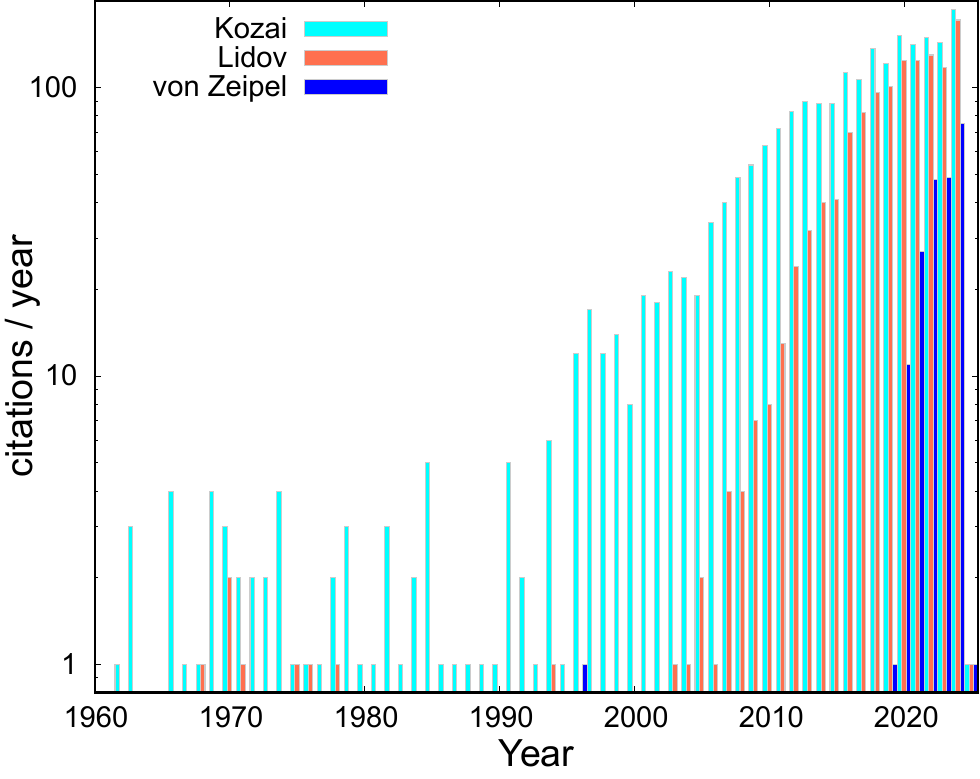}
\end{center}
\caption{%
Time series of the citation frequencies of
\citeauthor{kozai1962b}'s publications (cyan bars),
\citeauthor{lidov1962-en}'s  publications (red bars), and
\citeauthor{vonzeipel1910}'s publications (blue bars),
between 1961 and 2024 (and partly 2025) based on the ADS as of November 25, 2024.
Note that the statistics in 2025 at the bottom right part of the plot are incomplete because of the period we retrieved the citation records.
For the citation data for ``\citeauthor{kozai1962b},'' we bundle the citation frequencies of the two publications:
The main paper \citep{kozai1962b} and a meeting abstract by the same author with the same title published in the same issue of the same journal (abstracts of papers presented at the 111th Meeting of the American Astronomical Society at Yale University,  New Haven, Connecticut, August 26--29, 1962, on p. 579 of \textit{The Astronomical Journal,\/} \textit{67}, 1962).
Similarly, for the citation data for ``\citeauthor{lidov1961-en},'' we grouped the citation frequencies of closely relevant publications:
The original publication \citep{lidov1961-en},
its English translations \citep{lidov1962-en,lidov1963-en}, and
papers on the same topic \citep{lidov1963a-en,lidov1963b}.}
\label{fig:kozai-citation}
\end{figure}

All of the works by Kozai, Lidov, and von Zeipel deal with the same subject.
However, there is a significant difference in the paths these three works take toward recognition by the academic community.
As a prominent example highlighting the difference, in Figure \ref{fig:kozai-citation} we present the annual citation counts of their works since 1960 based on data stored on NASA's Astrophysics Data System (ADS).
Note that the vertical axis of Figure \ref{fig:kozai-citation} has a logarithmic scale.

As we mentioned, \citeauthor{lidov1961-en}'s achievement on doubly averaged CR3BP is equivalent to \citeauthor{kozai1962b}'s work.
However, Figure \ref{fig:kozai-citation} indicates that \citeauthor{lidov1961-en}'s work was not cited as frequently as \citeauthor{kozai1962b}'s.
We presume this is mainly due to the disparity in the popularity of the journals that published their papers, although Lidov's original paper written in Russian was quickly translated into English and published in two different journals \citep{lidov1962-en,lidov1963-en}.
To our knowledge, relying on ADS, the first publication citing both \citet{lidov1962-en} and \citet{kozai1962b} is \citet{lowrey1971}.
But for the next 33 years since \citet{lowrey1971}, it seems that not only the practical equivalence of these two works but \citeauthor{lidov1961-en}'s work has been largely forgotten.
Subsequently, after the publication of a paper on the secular dynamics of irregular satellites of giant planets \citep{cuk2004} that cites \citet{lidov1962-en} and \citet{kozai1962b} at the same time for the first time in the twenty-first century, \citeauthor{lidov1961-en}'s work began rapidly gaining attention.
Today, more and more people have come to know the equivalence of \citeauthor{lidov1961-en}'s work and \citeauthor{kozai1962b}'s work.

In contrast to Kozai's or Lidov's work, von Zeipel's work has been inconspicuous and perhaps even ostracized for a long time.
The only citation of \citet{vonzeipel1910} in the modern literature was made by \citet{bailey1996} until 2019.
However, \citeauthor{bailey1996}'s (\citeyear{bailey1996}) viewpoint was not on the phenomenon of the $e$--$I$ correlated oscillation or the conditions under which argument of pericenter $g$ of the perturbed body librates in the averaged CR3BP.
Since \citet{bailey1996} did not mention the whole picture of \citeauthor{vonzeipel1910}'s (\citeyear{vonzeipel1910}) work either, \citeauthor{vonzeipel1910}'s work continued to be buried even after \citeyear{bailey1996}.

Another reason may be that all of \citeauthor{vonzeipel1910}'s relevant publications (1898, 1901, 1910) were written in French.
It is no exaggeration to say that English is the only language accepted in the modern academic world.
Literature written in languages other than English is not considered, even if it may contain important findings.
\citet{ito2019} happened to be aware of the existence of the work by \citeauthor{vonzeipel1910} in the process of reading \citet{bailey1996}.
Then, \citeauthor{ito2019} scrutinized the contents of \citet{vonzeipel1910} with great interest.
As a result, \citeauthor{ito2019} found that the accuracy and completeness of \citeauthor{vonzeipel1910}'s series of works have the same merit as the later works by \citeauthor{lidov1961-en} and \citeauthor{kozai1962b}, and are even better in some aspects.
In other words, \citeauthor{vonzeipel1910}'s work is fully upward compatible with \citeauthor{lidov1961-en}'s and \citeauthor{kozai1962b}'s work.
Based on these facts,
we assert that the prefix ``von Zeipel--Lidov--Kozai'' should be used to designate the theoretical framework of this line of studies.

Among the three independent works, we summarize the unique aspects of Kozai's work as follows:
Kozai used electronic computers to perform a precise numerical quadrature of the disturbing function of several actual asteroids and discovered that (1373) Cincinnati's argument of pericenter $g$ will librate.
He continued his work along this line and later identified the locations of the stationary values of $(g,e)$ by drawing equipotential plots of the asteroid's disturbing Hamiltonian that are in typical mean motion resonance with Jupiter \citep[e.g.][he employed numerical quadrature here again]{kozai1985}.

At the end of this section, we'd like to note a new fact we found while preparing this paper.
First, \citet{hagihara1972-1} gave a detailed summary of \citet{kozai1962b} in his Section 9.25 ``Orbit with High Inclination and Eccentricity''.
Then, we found that \citet{hagihara1974-2} had pointed out the identity of Kozai's work on the doubly averaged inner CR3BP with what \citeauthor{vonzeipel1898} published at the end of the 19th century.
\citeauthor{hagihara1974-2} did not cite \citeauthor{vonzeipel1898}'s main work \citep{vonzeipel1910} but cited the previous ones \citep{vonzeipel1898,vonzeipel1901} in his book.
\citeauthor{hagihara1974-2} begins his Section ``14.6 von Zeipel's theory'' with the following statement:
\begin{quote}
``Consider with \citeauthor{vonzeipel1898} \citeyearpar{vonzeipel1898,vonzeipel1901} the three-body problem in which one of the masses is large compared with the other two. $[\cdots]$'' \citep[][his p. 538]{hagihara1974-2}
\end{quote}
\citeauthor{hagihara1974-2} further writes:
\begin{quote}
``The discussion, and hence the theory of Lindstedt, fails $[\cdots]$ if the orbital mutual inclination $I$ is greater than $39^\circ 14'$ for Jupiter and Saturn.
If the inclination is large enough, secular terms may appear that endanger the system's stability by the presence of linear terms in $t$.
% $[\cdots]$
The condition has been further studied by \citet{vonzeipel1901}, and recently by Kozai (Section 9.25).'' \citep[][his p. 550]{hagihara1974-2}
\end{quote}

The above statements indicate that \citeauthor{hagihara1974-2} was aware of von Zeipel's achievement in this line of study not long after the publication of \citet{kozai1962b}, including the threshold inclination value in the quadrupole level approximation: $I = 39^\circ 14'$.
This value is obviously derived from the theoretical framework described in Section \ref{ssec:Kozai-stationary} (see Eq. \eqref{eqn:I-range}).
Although it seems \citeauthor{hagihara1974-2} was not aware of the additional advanced work by \citet{vonzeipel1910}, the fact that \citeauthor{hagihara1974-2} had recognized \citeauthor{vonzeipel1910}'s early work already in the 1970s is quite impressive.

As far as we are aware, \citeauthor{hagihara1974-2}'s \citeyearpar{hagihara1974-2} recognition of \citeauthor{vonzeipel1901}'s early work did not receive subsequent attention from the broader academic community.
We browsed through \citet{hagihara1974-2} in the course of preparing our descriptions in Appendices \ref{appen:Pnodd} and \ref{appen:Wquad} when we happened to come across the above description.
\citeauthor{hagihara1974-2}'s series of treatises, including \citet{hagihara1974-2}, is not available in electronic form and is currently not easily accessible to the general public.
However, his works are so qualitatively detailed and quantitatively voluminous that they could be considered an encyclopedia of celestial mechanics.
By reviewing \citeauthor{hagihara1974-2}'s works and other historical works published in the early 20th century or the 19th century such as \citet{tisserand1889,tisserand1891,tisserand1894,tisserand1896} or \citet{poincare1892,poincare1893,poincare1899,poincare1905,poincare1907,poincare1909,poincare1910}, we believe that there is still a possibility of discovering new insights into the research history of the dynamical mechanisms corresponding to the von Zeipel--Lidov--Kozai oscillation.

\clearpage

\bibliographystyle{meep-pjab}

\bibliography{mybib-utf8}

\end{document}